\documentclass[10pt,letterpaper,twocolumn,english,aps,twocolumn]{revtex4-1}
\usepackage[T1]{fontenc}
\usepackage[latin9]{inputenc}
\usepackage{color}
\usepackage{babel}
\usepackage{array}
\usepackage{float}
\usepackage{multirow}
\usepackage{amsmath}
\usepackage{amssymb}
\usepackage{stackrel}
\usepackage{graphicx}
\usepackage[unicode=true,pdfusetitle,
 bookmarks=true,bookmarksnumbered=false,bookmarksopen=false,
 breaklinks=true,pdfborder={0 0 0},backref=false,colorlinks=true]
 {hyperref}
\hypersetup{
 pdfencoding=auto,linkcolor=blue,urlcolor=blue,citecolor=blue}
\usepackage{breakurl}

\makeatletter

\providecommand{\tabularnewline}{\\}

\usepackage{babel}
\usepackage{babel}

\makeatother

\begin{document}

\title{Permutation symmetry and entanglement in quantum states of heterogeneous
systems}

\author{Gururaj Kadiri}

\author{S. Sivakumar}

\email{gururaj@igcar.gov.in}
\email{siva@igcar.gov.in}

\affiliation{Materials Science Group, Indira Gandhi Centre for Atomic Research,
Kalpakkam, Tamilnadu, India. Pin: 603102}
\begin{abstract}
Permutation symmetries of multipartite quantum states are defined
only when the constituent subsystems are of equal dimensions. In this
work we extend this notion of permutation symmetry to heterogenous
systems, that is, systems composed of subsystems having unequal dimensions.
Given a tensor product space of $k$ subsystems (of arbitrary dimensions)
and a permutation operation $\sigma$ over $k$ symbols, these states
are such that they have identical decompositions (up to an overall
phase) in the given tensor product space and the tensor product space
obtained by the permuting the subsystems by $\sigma$. Towards this,
we construct a matrix whose action is to simultaneously permute the
subsystem label and subsystem dimension of a given state according
to permutation $\sigma$. Eigenvectors of this matrix have the required
symmetry. We then examine entanglement of states in the eigenspaces
of these matrices. It is found that all nonsymmetric eigenspaces of
such matrices are completely entangled subspaces, with states being
equally entangled in both the given tensor product space and the permuted
tensor product space. 
\end{abstract}


\maketitle

\section{Introduction}

Quantum theory is usually formulated in terms of states vectors which
are considered as elements of a suitable Hilbert space. For each classical
degree of freedom, the quantum formulation requires a corresponding
Hilbert space. Thus, a system of two 1D oscillators requires two Hilbert
spaces. If there are quantum degrees of freedom such as the spin of
a particle, they too will have their respective Hibert spaces. The
right way of describing the system with more than one degree of freedom
turns out to be the tensor product of the Hilbert spaces corresponding
to the various degrees of freedom relevant to the system, so such
tensor product spaces are central for the description of multipartite
quantum states. While pure states of multipartite quantum systems
could also be represented as a ray in $\mathbb{C}^{N}$ for an appropriate
$N$, the counterintutive features of such states, like the nonlocality,
entanglement etc, do not manifest in this unfactored space $\mathbb{C}^{N}$
but will manifest only in the tensor product of the Hilbert spaces
of the constituent systems.

A tensor product space (TPS) is homogeneous if the constituent subsystems
are of equal dimension. Otherwise, it is said to be heterogenous \cite{goyeneche2016multipartite}.
In the homogenous $k-$partite TPS having $d$ dimensional subsystems,
the ``symmetric subspace'' of $\mathbb{C}^{N}$(where $N=d^{k}$)
consists of states that remain invariant under arbitrary permutation
of their subsystem labels. The symmetric subspace is interesting because
its dimension scales with $k$ like the binomial coefficient $^{\left(d+k-1\right)}C_{k}$,
while the dimension of the composite system increases exponentially
like $d^{k}$. In the case of qubits ($d=2$), the symmetric subspace
is spanned by the Dicke basis \cite{dicke1954coherence}. Symmetric
states of homogeneous systems, particularly of the multipartite qubits,
have been extensively studied both experimentally \cite{Toth:07,PhysRevA.92.013629,Wei2015,PhysRevLett.106.130506,PhysRevLett.109.173604}
and theoretically \cite{markham2011entanglement,aulbach2012classification,PhysRevA.88.012305,PhysRevA.65.032328,2013arXiv1308.6595H},
with respect to their tomography \cite{PhysRevA.87.012109,PhysRevLett.105.250403,1367-2630-14-10-105001},
entanglement \cite{PhysRevA.67.022112,PhysRevA.78.052105,PhysRevA.81.054102,arnaud2016all}
etc. Though heterogeneous systems also have been studied theoretically
\cite{yu2008genuine,PhysRevA.69.012101,PhysRevA.74.052331,PhysRevA.87.062305,PhysRevA.88.062330,doi:10.1063/1.4790405},
and experimentally \cite{malik2016multi,xiao2014protecting}, the
notion of permutation symmetry is not readily extendable to them.
In this work, we demonstrate that there is a natural way to extend
the conventional notion of permutation symmetry to heterogeneous systems.

To motivate such a construction, consider a quantum system $S$, whose
Hilbert space is $H_{S}$ of dimension $d_{S}$. Assume that $S$
is allowed to interact with the environment $E$, whose Hilbert space
is $H_{E}$ of dimension $d_{E}$. The state of the composite system
$\left(S+E\right)$ can be represented in the tensor product space
$H_{S}\otimes H_{E}$ or the tensor product space $H_{E}\otimes H_{S}$.
Consider an arbitrary state $\left|\psi_{SE}\right\rangle $ in the
TPS $H_{S}\otimes H_{E}$:

\begin{equation}
\left|\psi_{SE}\right\rangle =\underset{i,j}{\sum}\alpha_{ij}\left|i\right\rangle \otimes\left|j\right\rangle 
\end{equation}
where $\left\{ \left|i\right\rangle \right\} _{i=0}^{d_{S}-1}$ and
$\left\{ \left|j\right\rangle \right\} _{j=0}^{d_{E}-1}$ are orthornomal
bases for the system and reservior respectively. The state ``physically
equivalent'' to $\left|\psi_{SE}\right\rangle $, in the TPS $H_{E}\otimes H_{S}$,
is

\begin{equation}
\left|\psi_{ES}\right\rangle =\underset{i,j}{\sum}\alpha_{ij}\left|j\right\rangle \otimes\left|i\right\rangle 
\end{equation}

State $\left|\psi_{ES}\right\rangle $ is physically equivalent to
$\left|\psi_{SE}\right\rangle $ in the sense that the expecation
value of any operator $\hat{M}$ of the system $S$ is identical in
both the states: $\left\langle \psi_{SE}|\hat{M}\otimes\hat{I}_{E}|\psi_{SE}\right\rangle =\left\langle \psi_{ES}|\hat{I}_{E}\otimes\hat{M}|\psi_{ES}\right\rangle $,
where $\hat{I}_{E}$ is the identity $H_{E}$. Similarly, reduced
density matrices corresponding to $S$, obtained by tracing out the
second subsystem from $\left|\psi_{SE}\right\rangle \left\langle \psi_{SE}\right|$
or the first subsystem from $\left|\psi_{ES}\right\rangle \left\langle \psi_{ES}\right|$
are identical. Further, the numerical measure of entanglement of the
state $\left|\psi_{SE}\right\rangle $ in the tensor product space
$H_{S}\otimes H_{E}$ is identical to that of the state $\left|\psi_{ES}\right\rangle $
in the tensor product space $H_{E}\otimes H_{S}$.

However, as tensor product operation is not commutative, $\left|j\right\rangle \otimes\left|i\right\rangle $
is not necessarily equal to $\left|i\right\rangle \otimes\left|j\right\rangle $,
and hence states $\left|\psi_{SE}\right\rangle $ and $\left|\psi_{ES}\right\rangle $
can be distinct when seen as states in $\mathbb{C}^{N}$. A state
$\left|\psi_{SE}\right\rangle $ is called exchange invariant if it
is identical to $\left|\psi_{ES}\right\rangle $, upto an overall
phase factor. In other words, a state $\left|\psi\right\rangle \in\mathbb{C}^{N}$
is exchange invaraint if it remains invariant under the transformation

\begin{equation}
\left|i\right\rangle \otimes\left|j\right\rangle \rightarrow\left|j\right\rangle \otimes\left|i\right\rangle 
\end{equation}
for all $i=0,\cdots d_{S}-1$ and $j=0,\cdots d_{E}-1$, where $\left\{ \left|i\right\rangle \right\} _{i=0}^{d_{S}-1}$
and $\left\{ \left|j\right\rangle \right\} _{j=0}^{d_{E}-1}$ are
two arbitrary orthonormal basis of the two subsystems.

For example, consider $d_{S}=2$ and $d_{E}=3$. Consider the computational
basis state $\left|3\right\rangle $ in $\mathbb{C}^{6}$. This state
in the $\mathbb{C}^{2}\otimes\mathbb{C}^{3}$ tensor product space
is $\left|1\right\rangle \otimes\left|0\right\rangle $. The physical
equivalent state of this in $\mathbb{C}^{3}\otimes\mathbb{C}^{2}$
is $\left|0\right\rangle \otimes\left|1\right\rangle $. But $\left|0\right\rangle \otimes\left|1\right\rangle $
in $\mathbb{C}^{3}\otimes\mathbb{C}^{2}$ corresponds to the state
$\left|1\right\rangle $ in $\mathbb{C}^{6}$, rather than $\left|3\right\rangle $
we began with. So state $\left|3\right\rangle $ is not symmetric
in the qubit-qutrit decomposition. Consider, on the other hand, the
computational basis state $\left|5\right\rangle $. This state in
$\mathbb{C}^{2}\otimes\mathbb{C}^{3}$ is $\left|1\right\rangle \otimes\left|2\right\rangle $.
The physical equivalent state of this in the $\mathbb{C}^{3}\otimes\mathbb{C}^{2}$
is $\left|2\right\rangle \otimes\left|1\right\rangle $. Since $\left|2\right\rangle \otimes\left|1\right\rangle $
in $\mathbb{C}^{3}\otimes\mathbb{C}^{2}$ corresponds to the same
state $\left|5\right\rangle $ in $\mathbb{C}^{6}$, state$\left|5\right\rangle $
is a symmetric state.

Similarly, consider the state $\frac{1}{\sqrt{3}}\left(\left|1\right\rangle +\left|2\right\rangle +\left|4\right\rangle \right)$
in $\mathbb{C}^{8}$. This state in the $\mathbb{C}^{2}\otimes\mathbb{C}^{4}$
is $\frac{1}{\sqrt{3}}\left(\left|01\right\rangle +\left|02\right\rangle +\left|10\right\rangle \right)$.
The physical equivalent state to this in the $\mathbb{C}^{4}\otimes\mathbb{C}^{2}$
is $\frac{1}{\sqrt{3}}\left(\left|10\right\rangle +\left|20\right\rangle +\left|01\right\rangle \right)$.
This state also corresponds to the same state $\frac{1}{\sqrt{3}}\left(\left|2\right\rangle +\left|4\right\rangle +\left|1\right\rangle \right)$
in $\mathbb{C}^{8}$, so $\frac{1}{\sqrt{3}}\left(\left|1\right\rangle +\left|2\right\rangle +\left|4\right\rangle \right)$
is a symmetric state in the qubit-ququart bipartite system.

This notion of exchange symmetry in heterogenous bipartite systems
can be extended to permutation symmetry of multipartite heterogenous
systems as well. First, notations to be used subsequently are explained.
A multiplicative partition of $N$ is represented by the $k-$tuple
$\mathbf{d}=\left[d_{1},d_{2},\cdots,d_{k}\right]$, where $d_{i}$s
are positive integers greater than $1$ such that $\underset{i}{\prod}d_{i}=N$.
The number of elements in $\mathbf{d}$ is denoted by $n\left(\mathbf{d}\right)$.
Corresponding to this $\mathbf{d}$, the $k-$partite TPS $\mathbb{C}^{d_{1}}\otimes\mathbb{C}^{d_{2}}\otimes\cdots\mathbb{\otimes C}^{d_{k}}$
is represented by $\mathbb{C^{\mathbf{d}}}$.

Let $\sigma$ be one of the elements of $S_{n\left(\mathbf{d}\right)}$,
the group of permutations over $n\left(\mathbf{d}\right)-$symbols.
Given $\mathbf{d}$ and a $\sigma$, another multiplicative partition
$\sigma\left(\mathbf{d}\right)$ of $N$ is obtained by permuting
the entries in $\mathbf{d}$ by $\sigma$, that is, $\sigma\left(\mathbf{d}\right)=\left[d_{\sigma^{-1}\left(1\right)},d_{\sigma^{-1}\left(2\right)},\cdots,d_{\sigma^{-1}\left(k\right)}\right]$.
The TPS corresponding to this partition is $\mathbb{C^{\sigma\left(\mathbf{d}\right)}}$.
As in the bipartite case, the state of a multipartite composite system
can be represented equally well in any of the TPS, $\mathbb{C^{\sigma\left(\mathbf{d}\right)}}$
for any $\sigma\in S_{n\mathbf{\left(d\right)}}$, although the number
of subsystems $n\mathbf{\left(d\right)}$ and the dimension $d_{i}$
of each subsystem are decided by the experiment.

Given the $k-$partite TPS $\mathbb{C^{\mathbf{d}}}$, a basis for
$\mathbb{C}^{N}$ is constructed from the tensor product of the $k$
bases $\mathbb{B}_{d_{1}},\mathbb{B}_{d_{2}},\cdots,\mathbb{B}_{d_{k}}$,
of the individual subsystems where $\mathbb{B}_{d_{r}}=\left\{ \left|i_{r}\right\rangle \right\} _{i_{r}=0}^{d_{r}-1}$
is an orthonormal basis for $\mathbb{C}^{d_{r}}$. This tensor product
basis is denoted by $\mathbb{B_{\mathbf{d}}}$. An element in $\mathbb{B_{\mathbf{d}}}$
is of the form $\left|i_{1}\right\rangle \otimes\left|i_{2}\right\rangle \otimes\cdots\otimes\left|i_{k}\right\rangle $,
where $\left|i_{r}\right\rangle \in\mathbb{B}_{d_{r}}$. This state
is expressed in short notation as $\left|i_{1}i_{2}\cdots i_{k}\right\rangle _{\mathbf{d}}$.

Similarly, another basis for $\mathbb{C}^{N}$ could be the tensor
product of the bases in the permuted order: $\mathbb{B}_{d_{\sigma^{-1}\left(1\right)}}\otimes\mathbb{B}_{d_{\sigma^{-1}\left(2\right)}}\otimes\cdots\otimes\mathbb{B}_{d_{\sigma^{-1}\left(k\right)}}$.
This basis is denoted as $\mathbb{B}_{\sigma\mathbf{\left(d\right)}}$,
suffix indicating that it has been obtained by a permutation of another
basis. An element in $\mathbb{B}_{\sigma\mathbf{\left(d\right)}}$
is of the form $\left|i_{\sigma^{-1}\left(1\right)}\right\rangle \otimes\left|i_{\sigma^{-1}\left(2\right)}\right\rangle \otimes\cdots\otimes\left|i_{\sigma^{-1}\left(k\right)}\right\rangle $
where $\left|i_{r}\right\rangle \in\mathbb{B}_{d_{r}}$. A short notation
for this state is as $\left|i_{\sigma^{-1}\left(1\right)}i_{\sigma^{-1}\left(2\right)}\cdots i_{\sigma^{-1}\left(k\right)}\right\rangle _{\sigma\left(\mathbf{d}\right)}$.

Given a TPS $\mathbf{\mathbb{C}^{\mathbf{d}}}$ and a permutation
$\sigma\in S_{n\left(\mathbf{d}\right)}$, a state $\left|\psi\right\rangle $
is invariant under permutation $\sigma$ if it remains invariant under
the mapping

\begin{equation}
\left|i_{1}i_{2}\cdots i_{k}\right\rangle _{\mathbf{d}}\rightarrow\left|i_{\sigma^{-1}\left(1\right)}i_{\sigma^{-1}\left(2\right)}\cdots i_{\sigma^{-1}\left(k\right)}\right\rangle _{\sigma\left(\mathbf{d}\right)},\label{eq:MP_Mapping}
\end{equation}
for all $0\leq i_{r}\leq d_{r}-1$ and $1\leq r\leq k$ where $k=n\left(\mathbf{d}\right)$.
In the bipartite case, $\sigma$ is the permutation $\left(1,2\right)$.

Towards achieving this mapping we construct an operator $\hat{T}_{\mathbf{d},\sigma}$:$\mathbb{B}_{\mathbf{d}}\rightarrow\mathbb{B}_{\sigma\mathbf{\left(d\right)}}$,
such that

\begin{equation}
\hat{T}_{\mathbf{d},\sigma}\left|i_{1}i_{2}\cdots i_{k}\right\rangle _{\mathbf{d}}=\left|i_{\sigma^{-1}\left(1\right)}i_{\sigma^{-1}\left(2\right)}\cdots i_{\sigma^{-1}\left(k\right)}\right\rangle _{\sigma\left(\mathbf{d}\right)}.\label{eq:T_MP_Action}
\end{equation}

Eigenvectors of $\hat{T}_{\mathbf{d},\sigma}$ are the states satisfying
the desired mapping defined in Eqn. \ref{eq:MP_Mapping}. Being a
a unitary transformation in $\mathbb{C}^{N}$, its eigenvalues are
complex numbers of unit modulii. Given a TPS $\mathbb{C^{\mathbf{d}}}$
and a permutation $\sigma$, the Hilbert space of the composite systems
$\mathbb{C}^{N}$ thus splits into disjoint eigenspaces of $\hat{T}_{\mathbf{d},\sigma}$:

\begin{equation}
\mathbb{C}^{N}\simeq\underset{\eta}{\bigoplus}\:\mathbb{S}_{\mathbf{d},\sigma}^{\eta}.\label{eq:Direct_Sum}
\end{equation}
Here $\mathbb{S}_{\mathbf{d},\sigma}^{\eta}$ is a subspace of $\mathbb{C}^{N}$,
composed of eigenstates of $\hat{T}_{\mathbf{d},\sigma}$ with eigenvalue
$\eta$. States in the subspaces $\mathbb{S}_{\mathbf{d},\sigma}^{\eta}$
are such that the reduced density matrix of the $r^{th}$ subsystem
in TPS $\mathbf{d}$ is identical to the reduced density matrix of
the $\sigma\left(r\right)^{th}$ subsystem in TPS $\sigma\left(\mathbf{d}\right)$.
This work provides a prescription for obtaining the dimensions and
bases of these subspaces.

The paper is organized as follows. A procedure for constructing bipartite
exchange invariant states is detailed in Section \ref{sec:Exchange_Symmetry}.
A multipartite extension of this construction to obtain states that
are invariant under an arbitrary permutation of subsystems is provided
in Section \ref{sec:Multipartite_Extension}. In Section \ref{sec:Permutation-symmetry-and_Entanglement},
we examine the entanglement of states in the subspaces $\mathbb{S}_{\mathbf{d},\sigma}^{\eta}$,
with respect to both the TPSs, $C^{\mathbf{d}}$ and $C^{\sigma\left(\mathbf{d}\right)}$.
In recent years, it has been argued that entanglement needs to be
defined with respect to a distinguished set of observables rather
than with respect to a distinguished tensor product space \cite{Zanardi2001,PhysRevLett.92.060402,viola2010entanglement,de2010entanglement,thirring2011entanglement}.
However, in this paper we stick to the conventional notion of entanglement,
but examine it in different tensor product spaces. Results are summarized
in Section \ref{sec:Summary}.

We provide a list of symbols appearing in this paper along with their
brief description in Tables \ref{tab:Bipartite_Symbols} and \ref{tab:MP_Symbols}.

\section{\label{sec:Exchange_Symmetry}Bipartite exchange symmetry}

In the bipartite case, the action of the matrix $\hat{T}_{\left[d_{1},d_{2}\right]}$
on the product state $\left|i\right\rangle \otimes\left|j\right\rangle \in\mathbb{B}_{d_{1}}\otimes\mathbb{B}_{d_{2}}$
is given by 
\begin{equation}
\hat{T}_{\left[d_{1},d_{2}\right]}\left(\left|i\right\rangle \otimes\left|j\right\rangle \right)=\left|j\right\rangle \otimes\left|i\right\rangle .\label{eq:T_action}
\end{equation}
The matrix representation of $\hat{T}_{\left[d_{1},d_{2}\right]}$
is the tensor commutator matrix (TCM) \cite{magnus1979commutation}.
The subscript $\left[d_{1},d_{2}\right]$ indicates that $\hat{T}_{\left[d_{1},d_{2}\right]}$
maps product states in $\mathbb{C}^{d_{1}}\otimes\mathbb{C}^{d_{2}}$
to the corresponding product states in $\mathbb{C}^{d_{2}}\otimes\mathbb{C}^{d_{1}}$.
The eigenvectors of $\hat{T}_{\left[d_{1},d_{2}\right]}$are the states
that are exchange invariant.

The explicit form of $\hat{T}_{\left[d_{1},d_{2}\right]}$ defined
as a mapping on the span of $\mathbb{B}_{d_{1}}\otimes\mathbb{B}_{d_{2}}$
is

\begin{equation}
\hat{T}_{[d_{1},d_{2}]}=\overset{d_{1}-1}{\underset{i=0}{\sum}}\overset{d_{2}-1}{\underset{j=0}{\sum}}\left(\left|j\right\rangle \otimes\left|i\right\rangle \right)\left(\left\langle i\right|\otimes\left\langle j\right|\right),\label{eq:T_Formula}
\end{equation}
where $\left\{ \left|i\right\rangle \right\} _{i=0}^{d_{1}-1}$ and
$\left\{ \left|j\right\rangle \right\} _{j=0}^{d_{2}-1}$ are arbitrary
bases for $\mathbb{C}^{d_{1}}$and $\mathbb{C}^{d_{2}}$ respectively.
In the computational basis of $\mathbb{C}^{N}$ the matrix elements
of $\hat{T}_{[d_{1},d_{2}]}$ are: 
\begin{eqnarray}
\left[\hat{T}_{\left[d_{1},d_{2}\right]}\right]_{m,n} & = & 1\mbox{ if }\begin{array}{c}
\left\lfloor \frac{m-1}{d_{1}}\right\rfloor =mod\left(n-1,d_{2}\right)\\
\mbox{and}\\
\left\lfloor \frac{n-1}{d_{2}}\right\rfloor =mod\left(m-1,d_{1}\right)
\end{array}\nonumber \\
 & = & 0,\mbox{ otherwise}.\label{eq:T_Matrix_Formula}
\end{eqnarray}
Here $1\leq m,n\leq N$ and $\left\lfloor x\right\rfloor $ denotes
the largest integer less than or equal to $x$. If $d_{1}=d_{2}=d$,
Eqn. \ref{eq:T_Formula} simplifies to the familiar permutation operator

\begin{equation}
\hat{T}_{[d,d]}=\overset{d-1}{\underset{i,j=0}{\sum}}\left|j\right\rangle \left\langle i\right|\otimes\left|i\right\rangle \left\langle j\right|.
\end{equation}
Exchange symmetric states are the eigenstates of this operator on
$\mathbb{C}^{N}$, $N=d_{1}d_{2}$.

For instance, if $d_{1}=d_{2}=2$, then

\begin{equation}
\hat{T}_{\left[2,2\right]}=\left[\begin{array}{cccc}
1 & 0 & 0 & 0\\
0 & 0 & 1 & 0\\
0 & 1 & 0 & 0\\
0 & 0 & 0 & 1
\end{array}\right].
\end{equation}
The eigenvalues of $\hat{T}_{\left[2,2\right]}$ are $\left\{ \pm1\right\} $,
with the symmetric subspace being three-dimensional spanned by $\left\{ \left|0\right\rangle ,\frac{1}{\sqrt{2}}\left(\left|1\right\rangle +\left|2\right\rangle \right),\left|3\right\rangle \right\} $,
which in $\mathbb{C}^{2}\otimes\mathbb{C}^{2}$ notation is $\left\{ \left|00\right\rangle ,\frac{1}{\sqrt{2}}\left(\left|01\right\rangle +\left|10\right\rangle \right),\left|11\right\rangle \right\} $.
The anti-symmetric subspace is one-dimensional, spanned by $\frac{1}{\sqrt{2}}\left(\left|1\right\rangle -\left|2\right\rangle \right)$
which in $\mathbb{C}^{2}\otimes\mathbb{C}^{2}$ is $\frac{1}{\sqrt{2}}\left(\left|01\right\rangle -\left|10\right\rangle \right)$,
the singlet Bell state.

Similarly, the matrix representation of $\hat{T}_{\left[2,3\right]}$
is

\begin{equation}
\hat{T}_{\left[2,3\right]}=\left[\begin{array}{cccccc}
1 & 0 & 0 & 0 & 0 & 0\\
0 & 0 & 1 & 0 & 0 & 0\\
0 & 0 & 0 & 0 & 1 & 0\\
0 & 1 & 0 & 0 & 0 & 0\\
0 & 0 & 0 & 1 & 0 & 0\\
0 & 0 & 0 & 0 & 0 & 1
\end{array}\right],
\end{equation}
whose eigenvalues are $\left\{ \pm1,\pm i\right\} $. The subspace
associated with eigenvalue $1$ is three-dimensional, 
\[
\mathbb{S}_{\left[2,3\right]}^{1}=\mbox{span}\left\{ \left|0\right\rangle ,\frac{1}{2}\left(\left|1\right\rangle +\left|2\right\rangle +\left|3\right\rangle +\left|4\right\rangle \right),\left|5\right\rangle \right\} ,
\]
where $\mathbb{S}_{\left[d_{1},d_{2}\right],\left(1,2\right)}^{\eta}$
is denoted by $\mathbb{S}_{\left[d_{1},d_{2}\right]}^{\eta}$. It
is easy to see that every vector in $\mathbb{S}_{\left[2,3\right]}^{1}$
is indeed exchange invariant. The subspace associated with eigenvalue
$-1$ is one-dimensional,

\begin{equation}
\mathbb{S}_{\left[2,3\right]}^{-1}=\mbox{span}\left\{ \frac{1}{2}\left(\left|1\right\rangle -\left|2\right\rangle -\left|3\right\rangle +\left|4\right\rangle \right)\right\} \label{eq:A_23}
\end{equation}

The eigenvectors of $\hat{T}_{\left[2,3\right]}$ have been expressed
in the basis for $\mathbb{C}^{6}$. To see their exchange symmetry,
the states are expressed in the $\mathbb{B}_{\left[2,3\right]}$ and
$\mathbb{B}_{\left[3,2\right]}$ bases. This requires to establish
a correspondence between the states in $\mathbb{B}$ and those in
$\mathbb{B}_{\left[d_{1},d_{2}\right]}$. Given one of the computational
basis states $\left|m\right\rangle $ in $\mathbb{B}_{N}$, its representation
in the tensor product basis $\mathbb{B}_{[d_{1},d_{2}]}$ is

\begin{equation}
\left|m\right\rangle =\left|i\right\rangle _{d_{1}}\otimes\left|j\right\rangle _{d_{2}}\equiv\left|i,j\right\rangle _{\left[d_{1},d_{2}\right]},\label{eq:Conversion_formula}
\end{equation}
where $i=\left\lfloor \frac{m}{d_{2}}\right\rfloor $, $j=mod\left(m,d_{2}\right)$.
Conversely, given a state $\left|i,j\right\rangle _{\left[d_{1},d_{2}\right]}\in\mathbb{C}^{d_{1}}\otimes\mathbb{C}^{d_{2}}$,
its representation in $\mathbb{C}^{N}$ is

\begin{equation}
\left|i,j\right\rangle _{\left[d_{1},d_{2}\right]}=\left|i\times d_{2}+j\right\rangle \label{eq:Back_Conversion_Formula}
\end{equation}

$\mathbb{S}_{\left[2,3\right]}^{-1}$ expressed in $\mathbb{C}^{2}\otimes\mathbb{C}^{3}$
is 
\[
\frac{1}{\sqrt{2}}\left(\left|0\right\rangle \otimes\frac{1}{\sqrt{2}}\left(\left|1\right\rangle -\left|2\right\rangle \right)-\left|1\right\rangle \otimes\frac{1}{\sqrt{2}}\left(\left|0\right\rangle -\left|1\right\rangle \right)\right),
\]
whereas in $\mathbb{C}^{3}\otimes\mathbb{C}^{2}$ this is $\frac{1}{\sqrt{2}}\left(\frac{1}{\sqrt{2}}\left(\left|0\right\rangle -\left|1\right\rangle \right)\otimes\left|1\right\rangle -\frac{1}{\sqrt{2}}\left(\left|1\right\rangle -\left|2\right\rangle \right)\otimes\left|0\right\rangle \right)$
which acquires an overall negative sign under simultaneous exchange
of subsystem states and dimensions. The respective reduced density
matrices are also identical,

\[
_{\left[2,3\right]}\rho_{1}={}_{\left[3,2\right]}\rho_{2}=\frac{1}{4}\left[\begin{array}{cc}
2 & 1\\
1 & 2
\end{array}\right],
\]
and

\[
_{\left[2,3\right]}\rho_{2}={}_{\left[3,2\right]}\rho_{1}=\frac{1}{4}\left[\begin{array}{ccc}
1 & -1 & 0\\
-1 & 2 & -1\\
0 & -1 & 1
\end{array}\right].
\]
where $_{\left[d_{1},d_{2}\right]}\rho_{i},\:i=1,2$ refers to the
reduced density matrix of the $i^{th}$ subsystem after tracing out
the other subsystem for a state $\left|\psi\right\rangle $ in the
$\mathbb{C}^{d_{1}}\otimes\mathbb{C}^{d_{2}}$ decomposition. The
corresponding reduced density matrices are identical for exchange
symmetric states. However, an arbitrary state $\left|\psi\right\rangle \in\mathbb{C}^{6}$
need not yield identical reduced density matrices as in this example

\begin{figure}[h]
\begin{centering}
\includegraphics[scale=0.3]{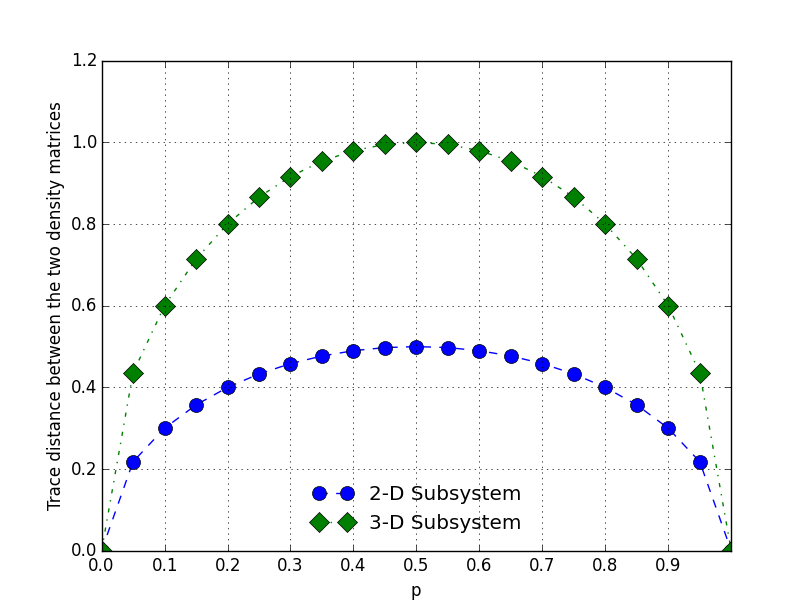} 
\par\end{centering}

\caption{Trace distances $d_{2}\left(p\right)$(blue plot) and $d_{3}\left(p\right)$
(green plot) as function of $p$, for $\left|\psi\left(p\right)\right\rangle $
given in Eqn. \ref{eq:Psi_p}.\label{fig:Trace-distance}}
\end{figure}

For instance, consider the state 
\begin{equation}
\left|\psi\left(p\right)\right\rangle =\sqrt{p}\left|0\right\rangle +\sqrt{1-p}\left(\frac{1}{2}\left(\left|1\right\rangle -\left|2\right\rangle -\left|3\right\rangle +\left|4\right\rangle \right)\right),\label{eq:Psi_p}
\end{equation}
which is a linear combination of one of the symmetric states $\left|0\right\rangle $
and anti-symmetric state of Eqn. \ref{eq:A_23}. This state is not
exchange symmetric unless $p=0,1$. Other values of $p$ correspond
to the state being asymmetric. The relevant reduced density matrices
of suitable dimensions are compared using trace distance. Denoting
the trace distance between the $2\times2$ density matrices $_{\left[2,3\right]}\rho{}_{1}$
and $_{\left[3,2\right]}\rho{}_{2}$ by $d_{2}\left(p\right)$ and
that between the $3\times3$ density matrices $_{\left[2,3\right]}\rho{}_{2}$
and $_{\left[3,2\right]}\rho{}_{1}$ by $d_{3}\left(p\right)$, we
have

\begin{eqnarray}
d_{2}\left(p\right) & = & \frac{1}{2}\underset{i}{\sum}\left|\lambda_{2,i}\right|,\;d_{3}\left(p\right)=\frac{1}{2}\underset{i}{\sum}\left|\lambda_{3,i}\right|\label{eq:Trace_distance}
\end{eqnarray}
where $\lambda_{2,i}$ are eigenvalues of $\left(_{\left[2,3\right]}\rho{}_{1}-{}_{\left[3,2\right]}\rho{}_{2}\right)$
and $\lambda_{3,i}$ are eigenvalues of $\left(_{\left[2,3\right]}\rho{}_{2}-{}_{\left[3,2\right]}\rho{}_{1}\right)$.
Figure \ref{fig:Trace-distance} shows the variation of $d_{2}\left(p\right)$
(blue plot) and $d_{3}\left(p\right)$ (green plot) as a function
of $p$.

The trace distances are symmetric about $p=1/2$, which corresponds
to the most asymmetric state. Any deviation from $p=1/2$ takes the
state $\left|\psi\left(p\right)\right\rangle $ closer to either symmetric
($p<1/2$) or antisymmetric ($p>1/2$) state. The trace distance peaks
at $p=1/2$, for which the $3\times3$ reduced density matrices $_{\left[2,3\right]}\rho{}_{2}$
and $_{\left[3,2\right]}\rho{}_{1}$ orthogonal to each-other:

\[
\begin{array}{cccccc}
_{\left[2,3\right]}\rho{}_{2} & = & \frac{1}{2}\left(\begin{array}{ccc}
1 & 0 & 0\\
0 & 0 & 0\\
0 & 0 & 1
\end{array}\right), & _{\left[3,2\right]}\rho{}_{1} & = & \left(\begin{array}{ccc}
0 & 0 & 0\\
0 & 1 & 0\\
0 & 0 & 0
\end{array}\right)\end{array}.
\]
It is to be noted that $_{\left[2,3\right]}\rho{}_{2}$ is a mixed
state whereas $_{\left[3,2\right]}\rho{}_{1}$ is a pure state. This
implies that the state $\left|\psi\left(p=0.5\right)\right\rangle $
is entangled in $\left[2,3\right]$ partition but separable in $\left[3,2\right]$
partition.

The other two eigenvectors of $\hat{T}_{\left[2,3\right]}$ also give
identical reduced density matrices in both the decompositions. One
marked difference between the case $\mathbf{d=}\left[2,2\right]$
discussed earlier and $\mathbf{d}=\left[2,3\right]$ case is the emergence
of eigenstates which acquire a phase $\neq0,\pi$ under subsystem
exchange operation. It will be demonstrated, for every heteogeneous
bipartite decomposition ($d_{1}\neq d_{2}$), there are subspaces
spanned by those states that acquire a phase $e^{i\phi},\phi\neq0,\pi$
under exchange of subsystems.

\subsection{$\hat{T}_{\left[d_{1},d_{2}\right]}$ as a permutation matrix }

Rules for relating the vectors in $\mathbb{B}$, and the TPS $\mathbb{B}_{\mathbf{d}}$
are already given in Eqs. \ref{eq:Conversion_formula} and \ref{eq:Back_Conversion_Formula}.
Vectors in the basis $\mathbb{B}_{\left[d_{1},d_{2}\right]}$ and
$\mathbb{B}_{\left[d_{2},d_{1}\right]}$ are related by the mapping
$\hat{T}_{\left[d_{1},d_{2}\right]}$, whose matrix representation
in the computational basis is a permutation matrix. Here, the permutation
effected by this matrix on the basis states is identified.

\begin{figure}[h]
\begin{centering}
\includegraphics[scale=0.2]{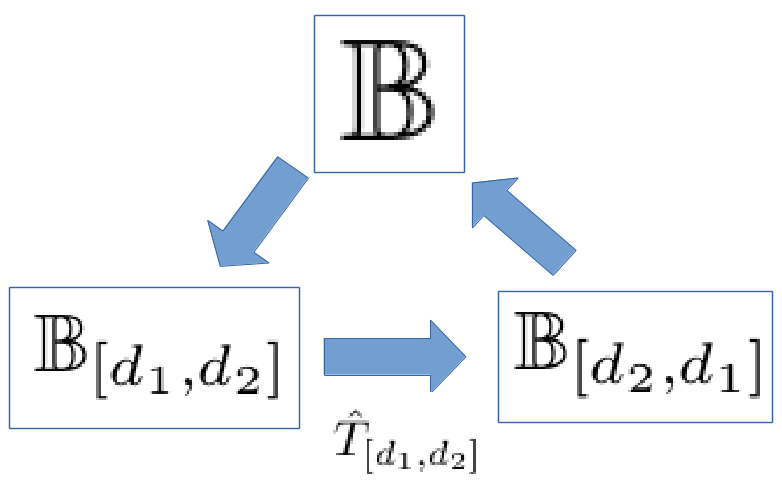} 
\par\end{centering}

\caption{\label{fig:Schematic_Cycle}Schematic of the procedure employed to
obtain the permutation corresponding to the permutation matrix $\hat{T}_{\left[d_{1},d_{2}\right]}$.}
\end{figure}

Towards this, begin with a state $\left|L_{n}\right\rangle \in\mathbb{B}$
where $0\leq L_{n}\leq d_{1}d_{2}-1$. Let the representation of this
state in $\mathbb{C}^{d_{1}}\otimes\mathbb{C}^{d_{2}}$ partition
be $\left|i,j\right\rangle _{[d_{1},d_{2}]}$ (refer Eqn. \ref{eq:Conversion_formula}).
The action of $\hat{T}_{\left[d_{1},d_{2}\right]}$ is to map this
state into the state $\left|j,i\right\rangle _{[d_{2},d_{1}]}\in\mathbb{C}^{d_{2}}\otimes\mathbb{C}^{d_{1}}$.
This corresponds to a state say $\left|L_{n+1}\right\rangle $ in
the unpartitioned space, where $L_{n+1}=j\times d_{1}+i$ (refer Eqn
\ref{eq:Back_Conversion_Formula}). See Fig. \ref{fig:Schematic_Cycle}
for the sequence of operations. This $\left|L_{n+1}\right\rangle $
can once again be expressed in the $\left[d_{1},d_{2}\right]$ partition
from which $L_{n+2}$ can be obtained by swapping the indices along
with the dimensions as was done for $L_{n+1}$. This process is repeated
until $L_{n+m+1}$ becomes $L_{n}$ for some $m$, which is guaranteed
since $\hat{T}_{\left[d_{1},d_{2}\right]}$ is one-to-one and, therefore,
invertible. States $\left(\left|L_{n}\right\rangle ,\left|L_{n+1}\right\rangle ,\cdots,\left|L_{n+m}\right\rangle \right)$
form one cycle. This cycle is called an $m-$cycle as it has $m$
states.

It is seen that index numbers $L_{n+1}$ can be obtained from $L_{n}$
using the relation:

\begin{equation}
L_{n+1}=d_{1}L_{n}-\left\lfloor \frac{L_{n}}{d_{2}}\right\rfloor \left(N-1\right).\label{eq:Guru_Formula}
\end{equation}

To generate another cycle, pickup another state from $\mathbb{B}$
not already present in any cycle as $\left|L_{n}\right\rangle $ and
generate another cycle in the same manner as above (or using Eqn.
\ref{eq:Guru_Formula}). Repeat the process until every state in $\mathbb{B}$
is accommodated in some cycle. It may be noted that the sum of the
lengths of the cycles equals the dimension of $\mathbb{B}$.

The action of $\hat{T}_{\left[d_{1},d_{2}\right]}$ is to group the
basis vectors of $\mathbb{B}$ into disjoint sets corresponding to
each cycle. The vectors in a given disjoint set are in the orbit of
the mapping $\hat{T}_{\left[d_{1},d_{2}\right]}$. Hence, this cycle
decomposition represents the permutation $\pi\left(d_{1},d_{2}\right)$
corresponding to the permutation matrix $\hat{T}_{\left[d_{1},d_{2}\right]}$.

As an illustration, $\pi\left(2,3\right)$ is explicitly constructed.
Here, $\mathbb{B}$ is $\left\{ \left|0\right\rangle ,\left|1\right\rangle ,\left|2\right\rangle ,\left|3\right\rangle ,\left|4\right\rangle \mbox{,\ensuremath{\left|5\right\rangle }}\right\} $.
Consider state $\left|0\right\rangle $ of $\mathbb{B}$. Its representation
in $\mathbb{C}^{2}\otimes\mathbb{C}^{3}$ is $\left|0\right\rangle \otimes\left|0\right\rangle $
which under the action of $\hat{T}_{\left[2,3\right]}$ goes over
to $\left|0\right\rangle \otimes\left|0\right\rangle $ in $\mathbb{C}^{3}\otimes\mathbb{C}^{2}$
which again corresponds to $\left|0\right\rangle $ in $\mathbb{C}^{6}$.
Similarly, $\left|1\right\rangle $ in $\mathbb{C}^{6}$ corresponds
to $\left|0\right\rangle \otimes\left|1\right\rangle $ in $\mathbb{C}^{2}\otimes\mathbb{C}^{3}$
which under the action of $\hat{T}_{\left[2,3\right]}$ goes to $\left|1\right\rangle \otimes\left|0\right\rangle $
in $\mathbb{C}^{3}\otimes\mathbb{C}^{2}$ which corresponds to $\left|2\right\rangle $
in $\mathbb{C}^{6}$. This is illustrated in Table \ref{tab:Bipartite_Example}.
One this is done, the cycles can be obtained easily: $\left|0\right\rangle $
in the left-most column is getting mapped to $\left|0\right\rangle $
in the right-most column, so $\left(0\right)$ is a $1-$cycle. Similarly
we have a sequence of states $\left|1\right\rangle \rightarrow\left|2\right\rangle \rightarrow\left|4\right\rangle \rightarrow\left|3\right\rangle \rightarrow\left|1\right\rangle $
so $\left(1,2,4,3\right)$ is another cycle. And $\left|5\right\rangle $
is another $1-$cycle. So the cycle decomposition corresponding to
$\hat{T}_{\left[2,3\right]}$ is $\pi\left(2,3\right)=\left(\left(0\right),\left(1,2,4,3\right),\left(5\right)\right)$.

\begin{table}[H]
\begin{centering}
\begin{tabular}{|c|c|c|c|c|c|c|}
\hline 
$\mathbb{B}$  & \multirow{7}{*}{$\Leftrightarrow$ } & $\mathbb{B}_{\left[2,3\right]}$  & \multirow{7}{*}{$\begin{array}{c}
\Rightarrow\\
\hat{T}_{\left[2,3\right]}\\
\Rightarrow
\end{array}$  } & $\mathbb{B}_{\left[3,2\right]}$  & \multirow{7}{*}{$\Leftrightarrow$  } & $\mathbb{B}$\tabularnewline
\cline{1-1} \cline{3-3} \cline{5-5} \cline{7-7} 
$\left|0\right\rangle $  &  & $\left|0\right\rangle \otimes\left|0\right\rangle $  &  & $\left|0\right\rangle \otimes\left|0\right\rangle $  &  & $\left|0\right\rangle $\tabularnewline
\cline{1-1} \cline{3-3} \cline{5-5} \cline{7-7} 
$\left|1\right\rangle $  &  & $\left|0\right\rangle \otimes\left|1\right\rangle $  &  & $\left|1\right\rangle \otimes\left|0\right\rangle $  &  & $\left|2\right\rangle $\tabularnewline
\cline{1-1} \cline{3-3} \cline{5-5} \cline{7-7} 
$\left|2\right\rangle $  &  & $\left|0\right\rangle \otimes\left|2\right\rangle $  &  & $\left|2\right\rangle \otimes\left|0\right\rangle $  &  & $\left|4\right\rangle $\tabularnewline
\cline{1-1} \cline{3-3} \cline{5-5} \cline{7-7} 
$\left|4\right\rangle $  &  & $\left|1\right\rangle \otimes\left|1\right\rangle $  &  & $\left|1\right\rangle \otimes\left|1\right\rangle $  &  & $\left|3\right\rangle $\tabularnewline
\cline{1-1} \cline{3-3} \cline{5-5} \cline{7-7} 
$\left|3\right\rangle $  &  & $\left|1\right\rangle \otimes\left|0\right\rangle $  &  & $\left|0\right\rangle \otimes\left|1\right\rangle $  &  & $\left|1\right\rangle $\tabularnewline
\cline{1-1} \cline{3-3} \cline{5-5} \cline{7-7} 
$\left|5\right\rangle $  &  & \multicolumn{1}{c||}{$\left|1\right\rangle \otimes\left|2\right\rangle $} &  & $\left|2\right\rangle \otimes\left|1\right\rangle $  &  & $\left|5\right\rangle $\tabularnewline
\hline 
\end{tabular}
\par\end{centering}

\caption{\label{tab:Bipartite_Example}Procedure for obtaining the action of
$\hat{T}_{\left[2,3\right]}$, and the cycle decomposition $\pi\left(2,3\right)$.}
\end{table}

Conventionally, $1-$cycles are not represented in the cycle decomposition
of a permutation. For clarity we shall include $1-$cycles also in
$\pi\left(d_{1},d_{2}\right)$. Cycle decomposition $\pi\left(d_{1},d_{2}\right)$
for some values of $d_{1}$ and $d_{2}$ are listed in Table \ref{tab:-Pi_Examples}
for illustration.

\begin{table}[h]
\begin{centering}
\begin{tabular}{|c|c|}
\hline 
$\left[d_{1},d_{2}\right]$  & $\pi\left(d_{1},d_{2}\right)$\tabularnewline
\hline 
\hline 
$\left[2,4\right]$  & $\left(\left(0\right),\left(1,2,4\right),\left(3,6,5\right),\left(7\right)\right)$\tabularnewline
\hline 
$\left[2,5\right]$  & $\left(\left(0\right),\left(1,2,4,8,7,5\right),\left(3,6\right),\left(9\right)\right)$\tabularnewline
\hline 
$\left[2,6\right]$  & $\left(\left(0\right),\left(1,2,4,8,5,10,9,7,3,6\right),\left(11\right)\right)$\tabularnewline
\hline 
$\left[3,3\right]$  & $\left(\left(0\right),\left(1,3\right),\left(2,6\right),\left(4\right),\left(5,7\right),\left(8\right)\right)$\tabularnewline
\hline 
$\left[3,4\right]$  & $\left(\left(0\right),\left(1,3,9,5,4\right),\left(2,6,7,10,8\right),\left(11\right)\right)$\tabularnewline
\hline 
$\left[3,5\right]$  & $\left(\left(0\right),\left(1,3,9,13,11,5\right),\left(2,6,4,12,8,10\right),\left(7\right),\left(14\right)\right)$\tabularnewline
\hline 
\end{tabular}
\par\end{centering}

\caption{$\pi\left(d_{1},d_{2}\right)$ for some values of $d_{1}$ and $d_{2}$
\label{tab:-Pi_Examples}}
\end{table}

\subsection{Eigenvectors and Eigenvalues of $\hat{T}_{\left[d_{1},d_{2}\right]}$}

Eigenvectors and eigenvalues of $\hat{T}_{\left[d_{1},d_{2}\right]}$
are obtained readily if the cycle $\pi\left(d_{1},d_{2}\right)$ is
known. Let the number of cycles be $p$ and and their respective cycle
lengths be $\left\{ l_{1},l_{2},\cdots,l_{p}\right\} $. Let $\omega_{l_{p}}$
be the primitive $l_{p}^{th}$ root of unity. Then the eigenvalues
of $\hat{T}_{\left[d_{1},d_{2}\right]}$ are given as \cite{stuart1991matrices}:

\begin{equation}
eig\left(T_{\left[d_{1},d_{2}\right]}\right)=\overset{p}{\underset{q=1}{\bigcup}}\left\{ \left(\omega_{l_{q}}\right)^{j}:1\leq j\leq l_{q}\right\} \label{eq:Eigen_Spectrum}
\end{equation}

Eigenvalues depend only on the lengths of the cycles in $\pi\left(d_{1},d_{2}\right)$
and not their elements. An $l-$cycle contributes $l$ eigenvalues
$\left\{ e^{\frac{2\pi i}{l}m},\mbox{ m }=0,\cdots,l-1\right\} $
to $\hat{T}_{\left[d_{1},d_{2}\right]}$.

Now, consider an $l-$cycle $\left(L_{1},L_{2},\cdots L_{l}\right)$.
Let one of the eigenvalues contributed by this cycle be $\lambda_{m}=e^{\frac{2\pi i}{l}m}$.
As $\hat{T}_{\left[d_{1},d_{2}\right]}$ generates the sequence $\left|L_{1}\right\rangle \rightarrow\left|L_{2}\right\rangle \rightarrow\cdots\rightarrow\left|L_{l}\right\rangle \rightarrow\left|L_{1}\right\rangle $,
it is easy to see that its action on a state $\left|\psi\right\rangle _{\lambda_{m}}$
will be equal to $\lambda_{m}\left|\psi\right\rangle _{\lambda_{m}}$if
$\left|\psi\right\rangle _{\lambda_{m}}$ is \cite{garcia2015eigenvectors}:
\begin{equation}
\left|\psi\right\rangle _{\lambda_{m}}=\frac{1}{\sqrt{l}}\left(\lambda_{m}^{-1}\left|L_{1}\right\rangle +\lambda_{m}^{-2}\left|L_{2}\right\rangle +\cdots+\lambda_{m}^{-l}\left|L_{l}\right\rangle \right)
\end{equation}
Eigenvectors corresponding to the same eigenvalue $\lambda_{m}$ span
a subspace called eigenspace. That the cycles are disjoint implies
that the eigenspaces corresponding to different eigvalues furnish
a direct sum decomposition of the composite Hilbert space, of the
form Eqn \ref{eq:Direct_Sum}. The eigenspace corresponding to the
eigenvalue $+1$, represented here as $\mathbb{S}_{\left[d_{1},d_{2}\right]}^{1}$,
is the symmetric subspace and that corresponding to $-1$, represented
here as $\mathbb{S}_{\left[d_{1},d_{2}\right]}^{-1}$, is the anti-symmetric
subspace. It is evident that there are as many symmetric eigenvectors
as there are cycles in $\pi\left(d_{1},d_{2}\right)$ and as many
anti-symmetric eigenvectors as the number of cycles of even length
in $\pi\left(d_{1},d_{2}\right)$.

For illustration, consider $d_{1}=2$ and $d_{2}=4$. Since $\pi\left(2,4\right)=\left(\left(0\right),\left(1,2,4\right),\left(3,6,5\right),\left(7\right)\right)$,
eigenvalues of $\hat{T}_{\left[2,4\right]}$ are $\left\{ 1,\omega,\omega^{2}\right\} $
where $\omega$ is the primitive cube root of unity, $e^{\frac{2\pi i}{3}}$.
The symmetric subspace $\mathbb{S}_{\left[2,4\right]}^{1}$ is four-dimensional,
given by: 
\[
\mbox{span}\left\{ \left|0\right\rangle ,\frac{1}{\sqrt{3}}\left(\left|1\right\rangle +\left|2\right\rangle +\left|4\right\rangle \right),\frac{1}{\sqrt{3}}\left(\left|3\right\rangle +\left|6\right\rangle +\left|5\right\rangle \right),\left|7\right\rangle \right\} 
\]
Similarly, eigenspaces corresponding to eigenvalues $\omega$ and
$\omega^{2}$, $\mathbb{S}_{\left[2,3\right]}^{\omega}$ and $\mathbb{S}_{\left[2,4\right]}^{\omega^{2}}$
are given by

\[
\mbox{span}\left\{ \frac{1}{\sqrt{3}}\left(\omega^{2}\left|1\right\rangle +\omega\left|2\right\rangle +\left|4\right\rangle \right),\frac{1}{\sqrt{3}}\left(\omega^{2}\left|3\right\rangle +\omega\left|5\right\rangle +\left|6\right\rangle \right)\right\} ,
\]
and

\[
\mbox{span}\left\{ \frac{1}{\sqrt{3}}\left(\omega\left|1\right\rangle +\omega^{2}\left|2\right\rangle +\left|4\right\rangle \right),\frac{1}{\sqrt{3}}\left(\omega\left|3\right\rangle +\omega^{2}\left|5\right\rangle +\left|6\right\rangle \right)\right\} 
\]
respectively. As there are no cycles of even length in $\pi\left(2,4\right)$,
there is no anti-symmetric subspace here.

When $d_{1}=d_{2}=d$, the matrix $\hat{T}_{\left[d,d\right]}$ is
involutory and its eigenvalues are $\left\{ \pm1\right\} $. Consequently,
there are only symmetric and anti-symmetric states. This feature is
brought out in $\pi\left(d,d\right)$ as well: it has $d$ fixed points
(cycles of unit length) corresponding to the states $\vert(d+1)i\rangle\in\mathbb{B}$,
$i=0\cdots d-1$ and the rest are 2-cycles. For example, $\pi\left(3,3\right)$
in Table \ref{tab:-Pi_Examples} is seen to have these features.

\begin{figure}[h]
\begin{centering}
\includegraphics[scale=0.33]{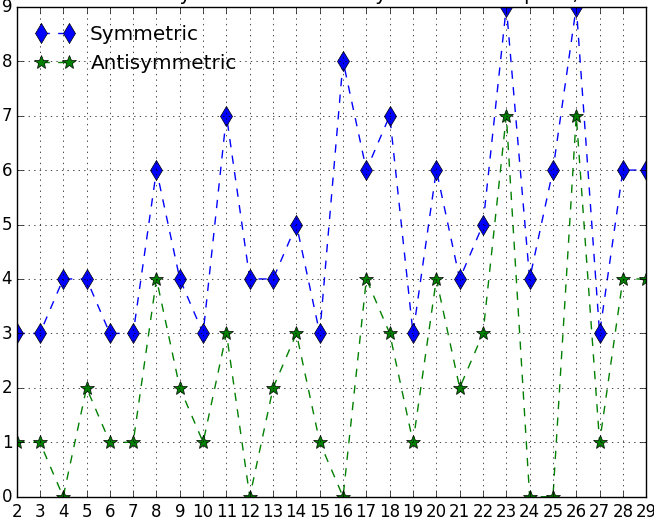} 
\par\end{centering}

\caption{Dimension of the symmetric (diamond) and anti-symmetric ( star) subspaces
of $T_{\left[2,d\right]}$, for different values of $d$.\label{fig:2_d2_symmetric_dimension}}
\end{figure}

The dimensions of symmetric and anti-symmetric subspaces do not necessarily
increase with increasing $N$ when $d_{1}\neq d_{2}$. This is also
evident from Table \ref{tab:-Pi_Examples} where the number of cycles
in $\pi\left(d_{1},d_{2}\right)$ does not necessarily increase with
$d_{1}$ or $d_{2}$. Figure \ref{fig:2_d2_symmetric_dimension} shows
the dimensions of the symmetric and anti-symmetric subspaces for $\hat{T}_{\left[2,d\right]}$
for $d=2$ to $29$.

The symmetric subspace of any $\hat{T}_{\left[d_{1},d_{2}\right]}$
is atleast three-dimensional, as any $\pi\left(d_{1},d_{2}\right)$
has atleast three cycles: two cycles corresponding to states $\left|0\right\rangle $
and $\left|N-1\right\rangle $ and another cycle comprising of the
rest of the states. Note that from Eqn. \ref{eq:Guru_Formula} it
follows that $\left(0\right)$ and $\left(N-1\right)$ are two fixed
points in $\pi\left(d_{1},d_{2}\right)$ for all $d_{1}$ and $d_{2}$.

Given $d_{1}$ and $d_{2}$, the eigenstates of $\hat{T}_{\left[d_{1},d_{2}\right]}$
furnish a special basis for $\mathbb{C}^{N}$. These basis vectors
have identical decompositions in $\mathbb{C}^{d_{1}}\otimes\mathbb{C}^{d_{2}}$
and $\mathbb{C}^{d_{2}}\otimes\mathbb{C}^{d_{2}}$ partitions. This
special basis is denoted by $\mathbb{B}_{\left[d_{1},d_{2}\right]}^{T}$.

From Eqn. \ref{eq:T_action}, it is seen that $\hat{T}_{[d_{1},d_{2}]}$
is the inverse of $\hat{T}_{\left[d_{2},d_{1}\right]}$. The eigenvalues
of $\hat{T}_{\left[d_{1},d_{2}\right]}$ come in complex conjugate
pairs. Hence $\hat{T}_{\left[d_{1},d_{2}\right]}$ and $\hat{T}_{\left[d_{2},d_{1}\right]}$
share the same set of eigenvalues. Further, the eigenspace corresponding
to an eigenvalue $\eta$ of $\hat{T}_{\left[d_{1},d_{2}\right]}$
will correspond to that of $\eta^{-1}$ for $\hat{T}_{\left[d_{2},d_{1}\right]}$
matrix and vice-versa:

\[
\mathbb{S}_{\left[d_{1},d_{2}\right]}^{\eta}=\mathbb{S}_{\left[d_{2},d_{1}\right]}^{\eta^{-1}}\forall~\mbox{eigenvalues \ensuremath{\eta} of }\hat{T}_{\left[d_{1},d_{2}\right]}
\]
This is reflected in the cycle decomposition also. By interchanging
$d_{1}$ and $d_{2}$ in Eqn. \ref{eq:Guru_Formula}, the cycles will
be generated in the reverse order so that $\pi\left(d_{2},d_{1}\right)=\pi^{-1}\left(d_{1},d_{2}\right)$.
For example, $\pi\left(2,4\right)=\left(\left(0\right),\left(1,2,4\right),\left(3,6,5\right),\left(7\right)\right)$
implies $\pi\left(4,2\right)=\left(\left(0\right),\left(1,4,2\right),\left(3,5,6\right),\left(7\right)\right)$.

If $m$ is the order of the cycle $\pi\left(d_{1},d_{2}\right)$,
then $\hat{T}_{\left[d_{1},d_{2}\right]}^{m}=\mathbb{I}$. More insights
about cycle structure of $\pi\left(d_{1},d_{2}\right)$ can be obtained
by examining the characteristic equation of $\hat{T}_{\left[d_{1},d_{2}\right]}$
derived in reference \cite{HENKDON1981135}. A cycle of length $l$
exists in $\pi\left(d_{1},d_{2}\right)$ if $l$ divides $l^{*}$,
where

\begin{equation}
l_{*}=min\left\{ p|d_{2}^{p}\equiv1\left(mod\:N-1\right)\right\} \label{eq:LStar-1}
\end{equation}

The number of cycles of length $l$ in $\pi\left(d_{1},d_{2}\right)$,
denoted by $\sigma\left(l\right)$, is given by

\begin{eqnarray}
\sigma(l) & = & gcd(d_{2}-1,N-1)+1\mbox{ if }l=1\nonumber \\
 & = & \underset{d|l}{\frac{1}{l}\sum}\mu\left(\frac{l}{d}\right)gcd\left(d_{2}^{d}-1,N-1\right)\mbox{ if }l>1\label{eq:N_l}
\end{eqnarray}
Here $\mu$ is the Mobius function defined on integers,

\begin{equation}
\mu(n)=\begin{cases}
1 & \mbox{if n=1,}\\
(-1)^{k} & \mbox{if \ensuremath{n} is the product of \ensuremath{k} distinct primes}\\
0 & \mbox{otherwise}
\end{cases}
\end{equation}
and $a|b$ stands for $a$ is divisor of $b$.

There is no anti-symmetric subspace when $l_{*}$ is odd, as there
are no cycles of even length, a consequence of the fact that the factors
of odd number are odd. For example, $l_{*}$ in the $\left[2,12\right]$
decomposition is $11$, so there is no anti-symmetric subspace for
$\hat{T}_{\left[2,12\right]}$ (see figure \ref{fig:2_d2_symmetric_dimension}).
Further, when $d_{1}=d_{2}=d$, $l_{*}\mbox{ is always }2$, giving
$\sigma\left(1\right)=d$ and $\sigma\left(2\right)=\frac{d\left(d-1\right)}{2}$,
so that symmetric subspace is $\frac{d\left(d+1\right)}{2}$ dimensional
and anti-symmetric subspace is $\frac{d\left(d-1\right)}{2}$ dimensional.

\section{\label{sec:Multipartite_Extension}Extension to multipartite qudit
states}

In the previous section, the notion of exchange symmetry for heterogeneous
bipartite systems has been discussed. In this section, the question
of generalizing this notion to multi-partite heterogeneous systems
is addressed.

\subsection{Multipartite subsystem permutation}

The first requirement is to identify a map between the computation
basis vectors of $\mathbb{B}$ of $\mathbb{C}^{N}$ to those in the
$\mathbb{B}_{\mathbf{d}}$ partition, akin to Eqn. \ref{eq:Conversion_formula}
for the bipartite case. Given any vector $\left|L\right\rangle $
in $\mathbb{B}$, its representation in $\mathbb{B}_{\mathbf{d}}$
partition is $\left|i_{1}i_{2}\cdots i_{k}\right\rangle _{\mathbf{d}}$
where each $i_{r}$ can be obtained successively as

\begin{equation}
i_{r}=\left\lfloor \frac{L-\overset{r-1}{\underset{j=1}{\sum}}i_{j}\left(\overset{k}{\underset{l=j+1}{\prod}}d_{l}\right)}{\overset{k}{\underset{m=r+1}{\prod}}d_{m}}\right\rfloor ,\label{eq:Conversion_Back}
\end{equation}
where $\left\lfloor x\right\rfloor $ represents the integer part
of $x$.

Conversely, given a basis state in $\left|i_{1}i_{2}\cdots i_{k}\right\rangle _{\mathbf{d}}\in\mathbb{B}_{\mathbf{d}}$,
the corresponding state in $\mathbb{B}$ is $\left|L\right\rangle $
where

\begin{equation}
L=\stackrel[r=1]{k}{\sum}i_{r}\times\left(\overset{k}{\underset{j=r+1}{\prod}}d_{j}\right).\label{eq:Conversion}
\end{equation}
Note that in Eqns. \ref{eq:Conversion_Back} and \ref{eq:Conversion},
the product of empty set is taken to be $1$.

\begin{figure}
\begin{centering}
\includegraphics[scale=0.25]{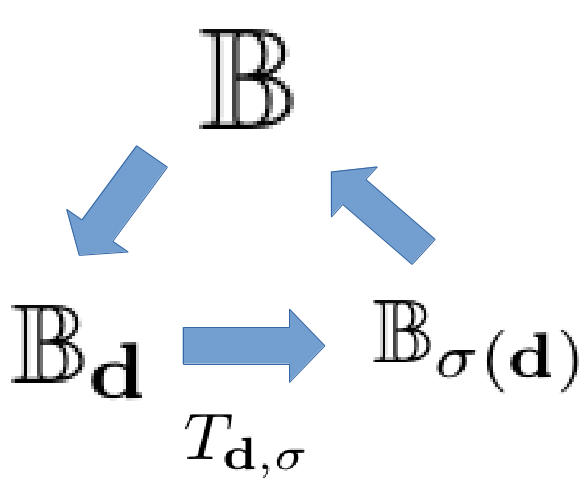} 
\par\end{centering}

\caption{\label{fig:Schematic_MP_Cycle}Schematic of the procedure employed
to obtain the permutation corresponding to the permutation matrix
$\hat{T}_{\mathbf{d},\sigma}$.}
\end{figure}

As in the bipartite case, the matrix representation of the mapping
$\hat{T}_{\mathbf{d},\sigma}$ in the computational basis of $\mathbb{C}^{N}$
yields a permutation matrix.

The schematic for this construction given in figure \ref{fig:Schematic_MP_Cycle},
is explained here. Start with one of the states $\left|L_{n}\right\rangle $
of $\mathbb{B}$, $0\leq L_{n}\leq N-1$. Its representation of in
$\mathbf{d}$ partition is $\left|i_{1}i_{2}\cdots i_{k}\right\rangle _{\mathbf{d}}$
which is obtained from Eqn. \ref{eq:Conversion_Back}. The action
of $\hat{T}_{\mathbf{d},\sigma}$ is to map this state to the state
$\left|i_{\sigma^{-1}\left(1\right)}i_{\sigma^{-1}\left(2\right)}\cdots i_{\sigma^{-1}\left(k\right)}\right\rangle _{\sigma\left(\mathbf{d}\right)}$.
Let $\left|L_{n+1}\right\rangle $ be the representation of this state
in the unpartitioned space, obtained through Eqn. \ref{eq:Conversion}
but using permuted $d_{j}$s. Now $\left|L_{n+1}\right\rangle $ is
expressed in $\mathbf{d}$ partition (using Eqn. \ref{eq:Conversion_Back}),
its labels and subsystems permuted and the new state is represented
again in the unpartitioned space using Eqn. \ref{eq:Conversion}.
Let that state be $\left|L_{n+2}\right\rangle $. This process is
repeated until $L_{n+m+1}$ becomes equal to $L_{n}$ for some $m$,
which is guaranteed since $\hat{T}_{\mathbf{d},\sigma}$ is one-to-one
and invertible. This way we have a mapping of states: $\left|L_{n}\right\rangle \rightarrow\left|L_{n+1}\right\rangle \rightarrow\cdots\rightarrow\left|L_{n+m+1}\right\rangle \rightarrow\left|L_{n}\right\rangle $,
which we shall indicate as an $m-$cycle $\left(L_{n},L_{n+1},\cdots,L_{n+m}\right)$.
To proceed, pick-up another vector from $\mathbb{\mathbb{B}}$ not
already appearing in cycles as $L_{n}$ and generate another cycle
of states as above. The process is repeated until every vector in
$\mathbb{B}$ is accommodated in some cycle. The permutation $\pi\in S_{N}$
corresponding to the permutation matrix $\hat{T}_{\mathbf{d},\sigma}$
is denoted by $\pi(\mathbf{d},\sigma)$.

To illustrate the scheme, $\pi\left(\left[2,2,3\right],\left(1,2,3\right)\right)$
is constructed. Towards this, begin with the computational basis $\mathbb{B}$
of $\mathbb{C}^{12}$. Consider one of the states of $\mathbb{B}$,
say $\left|1\right\rangle $. This state in $\left[2,2,3\right]$
is $\left|001\right\rangle _{\left[2,2,3\right]}$. Under $\left(1,2,3\right)$
permutation, this state goes over to the state $\left|100\right\rangle _{\left[3,2,2\right]}$
in $\left[3,2,2\right]$ decomposition, which corresponds to state
$\left|4\right\rangle $ in $\mathbb{B}$. Similarly, the mapping
for all the elements of $\mathbb{B}$ can be found as given in Table
\ref{tab:MP_Example}.

\begin{table}[H]
\begin{centering}
\begin{tabular}{|c|c|c|c|c|c|c|}
\hline 
$\mathbb{B}$  & \multirow{13}{*}{$\Leftrightarrow$ } & $\mathbb{B}_{\left[2,2,3\right]}$  & \multirow{13}{*}{$\begin{array}{c}
\Rightarrow\\
\hat{T}_{\left(\left[2,2,3\right],\left(1,2,3\right)\right)}\\
\Rightarrow
\end{array}$ } & $\mathbb{B}_{\left[3,2,2\right]}$  & \multirow{13}{*}{$\Leftrightarrow$ } & $\mathbb{B}$\tabularnewline
\cline{1-1} \cline{3-3} \cline{5-5} \cline{7-7} 
$\left|0\right\rangle $  &  & $\left|000\right\rangle $  &  & $\left|000\right\rangle $  &  & $\left|0\right\rangle $\tabularnewline
\cline{1-1} \cline{3-3} \cline{5-5} \cline{7-7} 
$\left|1\right\rangle $  &  & $\left|001\right\rangle $  &  & $\left|100\right\rangle $  &  & $\left|4\right\rangle $\tabularnewline
\cline{1-1} \cline{3-3} \cline{5-5} \cline{7-7} 
$\left|2\right\rangle $  &  & $\left|002\right\rangle $  &  & $\left|200\right\rangle $  &  & $\left|8\right\rangle $\tabularnewline
\cline{1-1} \cline{3-3} \cline{5-5} \cline{7-7} 
$\left|3\right\rangle $  &  & $\left|010\right\rangle $  &  & $\left|001\right\rangle $  &  & $\left|1\right\rangle $\tabularnewline
\cline{1-1} \cline{3-3} \cline{5-5} \cline{7-7} 
$\left|4\right\rangle $  &  & $\left|011\right\rangle $  &  & $\left|101\right\rangle $  &  & $\left|5\right\rangle $\tabularnewline
\cline{1-1} \cline{3-3} \cline{5-5} \cline{7-7} 
$\left|5\right\rangle $  &  & $\left|012\right\rangle $  &  & $\left|201\right\rangle $  &  & $\left|9\right\rangle $\tabularnewline
\cline{1-1} \cline{3-3} \cline{5-5} \cline{7-7} 
$\left|6\right\rangle $  &  & $\left|100\right\rangle $  &  & $\left|010\right\rangle $  &  & $\left|2\right\rangle $\tabularnewline
\cline{1-1} \cline{3-3} \cline{5-5} \cline{7-7} 
$\left|7\right\rangle $  &  & $\left|101\right\rangle $  &  & $\left|110\right\rangle $  &  & $\left|6\right\rangle $\tabularnewline
\cline{1-1} \cline{3-3} \cline{5-5} \cline{7-7} 
$\left|8\right\rangle $  &  & $\left|102\right\rangle $  &  & $\left|210\right\rangle $  &  & $\left|10\right\rangle $\tabularnewline
\cline{1-1} \cline{3-3} \cline{5-5} \cline{7-7} 
$\left|9\right\rangle $  &  & $\left|110\right\rangle $  &  & $\left|011\right\rangle $  &  & $\left|3\right\rangle $\tabularnewline
\cline{1-1} \cline{3-3} \cline{5-5} \cline{7-7} 
$\left|10\right\rangle $  &  & $\left|111\right\rangle $  &  & $\left|111\right\rangle $  &  & $\left|7\right\rangle $\tabularnewline
\cline{1-1} \cline{3-3} \cline{5-5} \cline{7-7} 
$\left|11\right\rangle $  &  & $\left|112\right\rangle $  &  & $\left|211\right\rangle $  &  & $\left|11\right\rangle $\tabularnewline
\hline 
\end{tabular}
\par\end{centering}

\caption{\label{tab:MP_Example}Procedure for obtaining the permutation $\pi\left(\left[2,2,3\right],\left(1,2,3\right)\right)$}
\end{table}

To generate the cycle, start with any element, say $\left|1\right\rangle $,
in the left-most column $\mathbb{B}$ . This state is mapped to state$\left|4\right\rangle $
of the right most column $\mathbb{B}$. State$\left|4\right\rangle $
on the left most column is mapped to$\left|5\right\rangle $ on the
right-most column and so on. This generates an orbit $\left|1\right\rangle \rightarrow\left|4\right\rangle \rightarrow\left|5\right\rangle \rightarrow\left|9\right\rangle \rightarrow\left|3\right\rangle \rightarrow\left|1\right\rangle $
so one of the cycles is $\left(1,4,5,9,3\right)$. Similarly starting
with $\left|2\right\rangle $ one generates another orbit$\left|2\right\rangle \rightarrow\left|8\right\rangle \rightarrow\left|10\right\rangle \rightarrow\left|4\right\rangle \rightarrow\left|7\right\rangle \rightarrow\left|6\right\rangle \rightarrow\left|2\right\rangle $,
yeilding a cycle $\left(2,8,10,4,7,6\right)$. Further, there are
two $1-$cycles, $\left(0\right)$ and $\left(11\right)$, so that
$\pi\left(\left[2,2,3\right],\left(1,2,3\right)\right)=\left(\left(0\right),\left(1,4,5,9,3\right),\left(2,8,10,4,7,6\right),\left(11\right)\right)$.

Using the same symbol $\pi$ for bipartite and multi-partite cases
should not lead to any confusion, as the arguments of $\pi$ are different
in the two cases. In fact, $\pi\left(d_{1},d_{2}\right)$ is a shorthand
notation for $\pi\left(\left[d_{1},d_{2}\right],\left(1,2\right)\right)$.
Also, note that $\sigma$ represents one of the permutations of the
subsystems, whereas $\pi$ refers to one of the permutations on the
computational basis vector labels: $\sigma\in S_{n\left(\mathbf{d}\right)}$
while $\pi\in S_{N}$. When $\sigma$ is the identity permutation
over $k-$symbols, $\sigma=\left(\left(1\right),\left(2\right),\cdots,\left(k\right)\right)$
where $k=n\left(\mathbf{d}\right)$, $\pi(\mathbf{d},\sigma)$ is
the identity permutation of $N$ symbols: $\pi(\mathbf{d},\sigma)=\left(\left(1\right),\left(2\right),\cdots,\left(N\right)\right)$.

For illustration, possible cycle decompositions $\pi\left(\left[2,2,3\right],\sigma\right)$
for all non-trivial permutations $\sigma\in S_{3}$ are given in Table
\ref{tab:Tripartite}.

\begin{table}[h]
\begin{centering}
\begin{tabular}{|c|c|}
\hline 
$\sigma$  & $\pi(\mathbf{d},\sigma)$\tabularnewline
\hline 
\hline 
$\left(\left(1,2\right),\left(3\right)\right)$  & $((0),(1),(2),(3,6),(4,7),(5,8),(9),(10),(11))$\tabularnewline
\hline 
$\left(\left(1,3\right),\left(2\right)\right)$  & $((0),(1,4,6),(2,8,9,3),(5,10,7),(11))$\tabularnewline
\hline 
$\left(\left(1\right),\left(2,3\right)\right)$  & $((0),(1,2,4,3),(5),(6),(7,8,10,9),(11))$\tabularnewline
\hline 
$\left(1,2,3\right)$  & $((0),(1,4,5,9,3),(2,8,10,7,6),(11))$\tabularnewline
\hline 
$\left(1,3,2\right)$  & $((0),(1,2,4,8,5,10,9,7,3,6),(11))$\tabularnewline
\hline 
\end{tabular}
\par\end{centering}

\caption{\label{tab:Tripartite}Permutation symmetries of $\left[2,2,3\right]$
decomposition of $N=12$.}
\end{table}

Once the cycle decomposition is available, obtaining eigenstates and
eigenvalues proceeds as in the bipartite case. For example, consider
the second entry of Table \ref{tab:Tripartite} corresponding to the
exchange of first and third subsystems in the $\left[2,2,3\right]$
decomposition. One of the cycles in $\pi\left(\left[2,2,3\right],\left(\left(1,3\right),\left(2\right)\right)\right)$
is $\left(1,4,6\right)$. It contributes three eigenstates to $\hat{T}_{\left[2,2,3\right],\left(\left(1,3\right),\left(2\right)\right)}$,
one of which is the symmetric state

\begin{eqnarray*}
\frac{1}{\sqrt{3}}\left(\left|1\right\rangle +\left|4\right\rangle +\left|6\right\rangle \right) & \equiv & \frac{1}{\sqrt{3}}\left(\left|001\right\rangle +\left|011\right\rangle +\left|100\right\rangle \right)_{\left[2,2,3\right]}\\
 & \equiv & \frac{1}{\sqrt{3}}\left(\left|001\right\rangle +\left|100\right\rangle +\left|110\right\rangle \right)_{\left[3,2,2\right]}.
\end{eqnarray*}

Many of the observations made in bipartite case hold in the multipartite
case as well. When $\sigma$ is transposition of two subsystems of
same dimensions, there are no cycles beyond two-cycles in $\pi\left(\mathbf{d},\sigma\right)$,
as exemplified by the entry corresponding to $\pi\left(\left[2,2,3\right],\left(\left(1,2\right),\left(3\right)\right)\right)$
in Table \ref{tab:Tripartite}. Next, the inverse of $\hat{T}_{\mathbf{d},\sigma}$
is not $\hat{T}_{\mathbf{d},\sigma^{-1}}$ but $\hat{T}_{\sigma\left(\mathbf{d}\right),\sigma^{-1}}$:

\begin{equation}
\hat{T}_{\mathbf{d},\sigma}=\hat{T}_{\sigma\left(\mathbf{d}\right),\sigma^{-1}}^{-1},\label{eq:T_T_Inv_Relation}
\end{equation}
from which it follows that $\pi\left(\sigma\left(\mathbf{d}\right),\sigma^{-1}\right)$
is the inverse of $\pi\left(\mathbf{d},\sigma\right)$ rather than
$\pi\left(\mathbf{d},\sigma^{-1}\right)$. Finally, given two permutations
$\sigma_{1},\sigma_{2}\in S_{n\left(\mathbf{d}\right)}$ the following
relation holds:

\begin{equation}
\pi(\mathbf{d},\sigma_{1}\circ\sigma_{2})=\pi(\sigma_{2}\left(\mathbf{d}\right),\sigma_{1})\circ\pi(\mathbf{d},\sigma_{2})\label{eq:Pi_Relation}
\end{equation}
where $\circ$ denotes the composition of permutations.

\subsection{Projection to the completely symmetric and antisymmetric subspaces}

For a homogenous $k-$partite partition $\mathbf{d}$, the completely
symmetric projector $\hat{S}_{\mathbf{d}}$ and completely antisymmetric
projector $\hat{A}_{\mathbf{d}}$ are defined as

\begin{equation}
\hat{S}_{\mathbf{d}}=\frac{1}{k!}\underset{\sigma}{\sum}\hat{T}_{\mathbf{d},\sigma},
\end{equation}
and 
\begin{equation}
\hat{A}_{\mathbf{d}}=\frac{1}{k!}\underset{\sigma}{\sum}\left(-1\right)^{sgn\left(\sigma\right)}\hat{T}_{\mathbf{d},\sigma},\label{eq:Antisymm_proj}
\end{equation}
where summation is over all $\sigma\in S_{k}$ (including the identity
element, for which $\hat{T}_{\mathbf{d},\sigma}$ is the identity
matrix) and $sgn\left(\sigma\right)$ is the parity of the permutation
$\sigma$. It is evident that in the homogenous case both $\hat{S}_{\mathbf{d}}$
and $\hat{A}_{\mathbf{d}}$ are projection operators, i.e., their
eigenvalues are $+1$ and $0$. Indeed, $^{\left(d+k-1\right)}C_{k}$
eigenvalues of $\hat{S}_{\mathbf{d}}$ are $+1$ and rest of them
are $0$. Similarly, $\hat{A}_{\mathbf{d}}$ has $^{d}C_{k}$ eigenvalues
as $+1$ and the other eigenvalues are $0$. If the eigenspaces of
these operators corresponding to eigenvalue $+1$ and $-1$ are $\mathbb{S}_{\mathbf{d}}$
and $\mathbb{A}_{\mathbf{d}}$ respectively, then

\begin{equation}
\left|\psi\right\rangle \in\mathbb{S}_{\mathbf{d}}\Longrightarrow\hat{T}_{\mathbf{d},\sigma}\left|\psi\right\rangle =\left|\psi\right\rangle ,\:\forall\sigma\in S_{n\left(\mathbf{d}\right)},
\end{equation}

and

\begin{equation}
\left|\psi\right\rangle \in\mathbb{A}_{\mathbf{d}}\Longrightarrow\hat{T}_{\mathbf{d},\sigma}\left|\psi\right\rangle =\left(-1\right)^{sgn\left(\sigma\right)}\left|\psi\right\rangle ,\:\forall\sigma\in S_{n\left(\mathbf{d}\right)}.
\end{equation}
When $k>d$, $\hat{A}_{\mathbf{d}}$ is a zero matrix and there is
no completely antisymmetric subspace in that case \cite{arnaud2016all}.

From the projectors $\hat{S}_{\mathbf{d}}$ and $\hat{A}_{\mathbf{d}}$,
two (mixed) states $\rho_{S}$ and $\rho_{A}$ are defined to be 
\begin{eqnarray}
\rho_{S} & = & \frac{1}{^{\left(d+k-1\right)}C_{k}}\hat{S}_{\mathbf{d}},\nonumber \\
\rho_{A} & = & \frac{1}{^{d}C_{k}}\hat{A}_{\mathbf{d}}.\label{eq:Mixed_States}
\end{eqnarray}
The density matrix $\rho_{A}$ is called the ``antisymmetric state''
and its entanglement is studied in \cite{Christandl2012}. $\rho_{A}$
is found to be maximally steerable for all dimensions \cite{PhysRevLett.112.180404}.
A one parameter family of states is constructed using these states
as 
\begin{equation}
\rho\left(p\right)=p\rho_{S}+\left(1-p\right)\rho_{A},
\end{equation}
where $p\in\left[0,1\right]$. The states $\rho\left(p\right)$ are
such that they remain invariant under any local unitary transformation
acting identically on all the subsystems:

\begin{equation}
\rho\left(p\right)=\left(\underset{\mbox{k times}}{\underbrace{U_{d}\otimes U_{d}\cdots\otimes U_{d}}}\right)\rho\left(p\right)\left(\underset{\mbox{k times}}{\underbrace{U_{d}^{\dagger}\otimes U_{d}^{\dagger}\cdots\otimes U_{d}^{\dagger}}}\right)
\end{equation}
where $U_{d}$ is a $d\times d$ unitary matrix. In the bipartite
setting, $k=2$, $\rho\left(p\right)$ are the well-known Werner states
\cite{PhysRevA.40.4277}. Separability of the these states in the
tripartite case is discussed in \cite{PhysRevA.63.042111}.

For heterogenous $\mathbf{d}$, $\hat{S}_{\mathbf{d}}$ is no longer
a projector since some of its eigenvalues are different from $0$
and $1$. Eigenspace of $\hat{S}_{\mathbf{d}}$ corresponding to an
eigenvalue $+1$ is three-dimensional, with three eigenvectors being
$\left|0\right\rangle $, $\left|N-1\right\rangle $ and $\left|\Gamma_{N}\right\rangle $,
for all $\mathbf{d}\in S_{N}$, where $\left|\Gamma_{N}\right\rangle $
is defined as

\begin{equation}
\left|\Gamma_{N}\right\rangle =\frac{1}{\sqrt{N-2}}\left(\stackrel[n=1]{N-2}{\sum}\left|n\right\rangle \right),\label{eq:Gamma_State}
\end{equation}
where $\left|n\right\rangle $ is the $\left(n+1\right)^{th}$ computational
basis for $\mathbb{C}^{N}$. We refer to the subspace spanned by the
three vectors $\left|0\right\rangle $, $\left|N-1\right\rangle $
and $\left|\Gamma_{N}\right\rangle $ as the ``generalized symmetric
subspace'', as it belongs to the symmetric subspace corresponding
to any permutation in any TPS:

\begin{equation}
\text{span}\left\{ \left|0\right\rangle ,\left|N-1\right\rangle ,\left|\Gamma_{N}\right\rangle \right\} \subseteq\mathbb{S}_{\mathbf{d},\sigma}^{1}\forall\mathbf{d}\in\mathbb{P}\left(N\right),\sigma\in S_{n\left(\boldsymbol{d}\right)}\label{eq:GSS}
\end{equation}
Similarly, $\hat{A}_{\mathbf{d}}$ is not a projection operator in
the heterogenous case and its eigenvalues are of magnitude strictly
less than one.

\subsection{Equivalent decompositions and Coarse-graining}

Given a state in $\mathbb{C}^{N}$, there could be different tensor
product spaces $\mathbb{C}^{\mathbf{d}_{1}}$ and $\mathbb{C}^{\mathbf{d}_{2}}$
consistent with $N$, but $\mathbf{d}_{1}$ and $\mathbf{d}_{2}$
not related by any permutation symmetry. For example, $N=8$ may be
realized in two ways: $\mathbf{d}_{1}=\left[2,2,2\right]$ and $\mathbf{d}_{2}=\left[2,4\right]$.
How are the cycle decompsotions $\pi\left(\mathbf{d}_{1},\sigma\right)$
and $\pi\left(\mathbf{d}_{2},\sigma\right)$ related?

We represent all the multiplicative partitions of $N$ (including
those that differ in the order of subsystems) as $\mathbb{P}\left(N\right)$:

\begin{equation}
\mathbb{P}\left(N\right)=\left\{ \boldsymbol{d}:\prod d_{i}=N\right\} 
\end{equation}
For example, $N=12$ allows for the following seven multiplicative
partitions: 
\[
\mathbb{P}\left(12\right)=\left\{ \left[2,2,3\right],\left[2,3,2\right],\left[2,6\right],\left[3,2,2\right],\left[3,4\right],\left[4,3\right],\left[6,2\right]\right\} .
\]

Among these, let $\mathbb{P}_{k}\left(N\right)$ denote the set of
all partitions $\mathbf{d}\in\mathbb{P}\left(N\right)$ having $n\left(\mathbf{d}\right)=k$.
For example, 
\begin{eqnarray*}
\mathbb{P}_{2}\left(12\right) & = & \left\{ \left[2,6\right],\left[3,4\right],\left[4,3\right],\left[6,2\right]\right\} 
\end{eqnarray*}
and

\[
\mathbb{P}_{3}\left(12\right)=\left\{ \left[2,2,3\right],\left[2,3,2\right],\left[3,2,2\right]\right\} .
\]
The largest value of $n\left(\mathbf{d}\right)$ is equal to $\Omega\left(N\right)$,
the number of prime factors of $N$ (allowing for repetitions). Given
a partition $\mathbf{d}$ and a permutation $\sigma\in S_{n\left(\mathbf{d}\right)}$,
define $\sigma\left(\mathbf{d}\right)$ as the $k-$tuple $\left[d_{\sigma^{-1}\left(1\right)},d_{\sigma^{-1}\left(2\right)},\cdots,d_{\sigma^{-1}\left(k\right)}\right]$.
Further, we denote the equivalence class (under permutation) of set
of all decompositions connected to a partition $\boldsymbol{d}_{e}$
by $\mathbb{E}$

\begin{equation}
\mathbb{E}\left(\boldsymbol{d}_{e}\right)=\left\{ \boldsymbol{d}'\in\mathbb{P}_{n\left(\mathbf{d}\right)}\left(N\right)|\exists\sigma\in S_{n\left(\mathbf{d}\right)},\boldsymbol{d'}=\sigma\left(\boldsymbol{d}_{e}\right)\right\} 
\end{equation}

For example, in case of $N=12$, we have three distinct classes

\begin{eqnarray*}
\mathbb{E}\left(\left[2,6\right]\right) & = & \left\{ \left[2,6\right],\left[6,2\right]\right\} \\
\mathbb{E}\left(\left[3,4\right]\right) & = & \left\{ \left[3,4\right],\left[4,3\right]\right\} \\
\mathbb{E}\left(\left[2,2,3\right]\right) & = & \left\{ \left[2,2,3\right],\left[2,3,2\right],\left[3,2,2\right]\right\} 
\end{eqnarray*}
Here an equivalence class is labeled by one of its members $\boldsymbol{d}_{e}$
whose entries are arranged in increasing order: $d_{i}\leq d_{j}$,
if $i<j$. We call $\boldsymbol{d}_{e}$ a representative partition
of the class to which it belongs.

Among the representative partitions of $N$, we identify one which
contains only prime $d_{i}$s. We call this, the ``primitive decomposition''
and represent it by $\boldsymbol{d}_{p}$. For example, for $N=24$,
the primitive partition is $\mathbf{d}_{p}=\left[2,2,2,3\right]$.
Further, we call partitions $\boldsymbol{d}\in\mathbb{E}\left(\boldsymbol{d}_{p}\right)$,
the prime partitions. By the uniqueness of prime factorization we
have $\mathbb{E}\left(\boldsymbol{d}_{p}\right)=\mathbb{P}_{\Omega\left(N\right)}\left(N\right)$.

If the cycle-decomposition $\pi\left(\mathbf{d}_{e},\sigma\right)$
of the representative partitions $\mathbf{d}_{e}$ is obtained for
all $\sigma\in S_{n\left(\mathbf{d}_{e}\right)}$, the decompositions
$\pi\left(\mathbf{d},\sigma_{2}\right)$, corresponding to any other
partition $\mathbf{d}$ belonging to the same class $\mathbb{E}\left(\mathbf{d}_{e}\right)$
can be obtained. Permutations $\pi\left(\mathbf{d},\sigma_{2}\right)$,
for $\mathbf{d}\in\mathbb{E}\left(\boldsymbol{d}_{e}\right)$ and
$\sigma_{2}\in S_{n\left(\mathbf{d}_{e}\right)}$, can be obtained
from the permutations corresponding to the representative partition
$\mathbf{d}_{e}$ through the relation

\begin{equation}
\pi\left(\mathbf{d},\sigma_{2}\right)=\pi\left(\sigma_{1}\left(\mathbf{d}_{e}\right),\sigma_{2}\right)=\pi\left(\mathbf{d}_{e},\sigma_{2}\circ\sigma_{1}\right)\circ\pi^{-1}\left(\mathbf{d}_{e},\sigma_{1}\right)
\end{equation}
This relation is obtained by just rearranging the Eqn. \ref{eq:Pi_Relation}.
As $\sigma_{2}\circ\sigma_{1}$ is another permutation belonging to
$S_{n\left(\mathbf{d}_{e}\right)}$, it follows that permutation symmetries
of every tensor product space can be obtained using the permutation
symmetries of representative decomposition $\mathbf{d}_{e}$ alone.

Now, consider a TPS $\mathbf{\mathbb{C}}^{\boldsymbol{d}^{\prime}}$,
where $\boldsymbol{d}^{\prime}\in\mathbb{P}_{k^{\prime}}\left(N\right)$
for $k^{\prime}<\Omega\left(N\right)$. Such partitions with fewer
number of subsystems than the prime partition are called coarse-grained
partitions. It is important to know whether the permutations $\pi\left(\mathbf{d}^{\prime},\sigma^{\prime}\right)$
of the coarse grained partitions are related to those of the primitive
decomposition, $\pi\left(\mathbf{d}_{p},\sigma\right)$.

As a coarse-grained partition $\boldsymbol{d}^{\prime}$ involves
a fewer number of tensor products to generate $\mathbb{C}^{N}$ than
the maximal number of tensor products $\Omega_{N}$ in $\mathbf{d}_{p}$.
Therefore the coarse-grained partition can be expressed by combining
(via tensoring) some of the prime dimensional Hilbert spaces. Each
of the dimensions $d_{r}^{\prime}$ in $\mathbf{d}^{\prime}$ is a
product of one or more $d_{i}$'s of $\mathbf{d}_{p}$. Hence, the
cycle decomposition $\pi\left(\mathbf{d}^{\prime},\sigma^{\prime}\right)$
is identical to that of $\pi\left(\mathbf{d},\sigma_{2}\right)$ for
some $\mathbf{d}\in\mathbb{E}\left(\mathbf{d}_{p}\right)$ and $\sigma_{2}\in S_{n\left(\mathbf{d}_{p}\right)}$.
In essence, given $\pi\left(\mathbf{d}^{\prime},\sigma^{\prime}\right)$,
it is always possible to find two permutations $\sigma_{1},\sigma_{2}\in S_{n\left(\mathbf{d}_{p}\right)}$
such that

\begin{equation}
\pi\left(\mathbf{d}^{\prime},\sigma^{\prime}\right)=\pi\left(\sigma_{1}\left(\mathbf{d}_{p}\right),\sigma_{2}\right)\label{eq:Coarse_Graining_Cycle}
\end{equation}

For example, consider $N=24$. Its primitive decomposition is $\mathbf{d}_{p}=\left[2,2,2,3\right]$.
Consider a coarse-grained decomposition of $N$, say, $\mathbf{d}^{\prime}=\left[4,3,2\right]$
and the permutation operation to be the anti-cyclic rotation $\sigma^{'}=\left(1,3,2\right)$.
In this case, $\pi\left(\mathbf{d}^{\prime},\sigma^{\prime}\right)$
is: 
\[
\pi\left(\mathbf{d}^{\prime},\sigma^{\prime}\right)=\left(\begin{array}{c}
\left(0\right),\\
\left(1,4,16,18,3,12,2,8,9,13,6\right),\\
\left(5,20,11,21,15,14,10,17,22,19,7\right),\\
\left(23\right).
\end{array}\right).
\]
Permutations $\sigma_{1},\sigma_{2}\in S_{4}$ such that $\pi\left(\sigma_{1}\left(\mathbf{d}_{p}\right),\sigma_{2}\right)=\pi\left(\left[4,3,2\right],\left(1,3,2\right)\right)$
is $\sigma_{1}=\left(\left(1\right),\left(2\right),\left(3,4\right)\right)\mbox{ and }\sigma_{2}=\left(\left(1,3\right),\left(2,4\right)\right)$.

If attention is restricted to bipartite partitioning $\mathbf{d}^{'}=\left[d_{1}^{'},d_{2}^{'}\right]$,
where the only non-trivial permutation is the subsystem exchange $\sigma'=\left(2,1\right)$,
it is possible to find suitabe $\sigma_{1},\sigma_{2}\in S_{\Omega\left(N\right)}$
such that $\pi\left(\sigma_{1}\left(\mathbf{d}_{p}\right),\sigma_{2}\right)=\pi\left(d_{1}^{\prime},d_{2}^{\prime}\right)$
where $\sigma\left(d_{1}^{\prime},d_{2}^{\prime}\right)$ is the permutation
corresponding to the bipartite exchange. This is illustrated with
an example. If $N=24$, the allowed bipartite partitions are 
\[
\mathbb{P}_{2}\left(24\right)=\left\{ \left[2,12\right],\left[3,8\right],\left[4,6\right],\left[6,4\right],\left[8,3\right],\left[12,2\right]\right\} 
\]
The primitive decomposition $\mathbf{d}_{p}$ for $N=24$ is $\mathbf{d}_{p}=\left[2,2,2,3\right]$.
For every $\mathbf{d^{\prime}\in}\mathbb{P}_{2}\left(24\right)$,
Table \ref{tab:-bipartition_24} shows possible $\sigma_{1},\sigma_{2}\in S_{4}$
satisfying Eqn. \ref{eq:Coarse_Graining_Cycle}, that is $\pi\left(\sigma_{1}\left(\left[2,2,2,3\right]\right),\sigma_{2}\right)=\pi\left(\mathbf{d}^{\prime},\left(1,2\right)\right)$.

\begin{table}[h]
\begin{centering}
\begin{tabular}{|c|c|c|c|}
\hline 
$\mathbf{d}^{'}$  & $\sigma_{1}$  & $\sigma_{1}\left(\mathbf{d}_{p}\right)$  & $\sigma_{2}$\tabularnewline
\hline 
\hline 
$\left[2,12\right]$  & $\left(\left(1\right),\left(2\right),\left(3\right),\left(4\right)\right)$  & $\left[2,2,2,3\right]$  & $\left(1,4,3,2\right)$\tabularnewline
\hline 
$\left[3,8\right]$  & $\left(1,2,3,4\right)$  & $\left[3,2,2,2\right]$  & $\left(1,4,3,2\right)$\tabularnewline
\hline 
$\left[4,6\right]$  & $\left(\left(1\right),\left(2\right),\left(3,4\right)\right)$  & $\left[2,2,3,2\right]$  & $\left(\left(1,3\right),\left(2,4\right)\right)$\tabularnewline
\hline 
$\left[6,4\right]$  & $\left(\left(1\right),\left(2,3,4\right)\right)$  & $\left[2,3,2,2\right]$  & $\left(\left(1,3\right),\left(2,4\right)\right)$\tabularnewline
\hline 
$\left[8,3\right]$  & $\left(\left(1\right),\left(2\right),\left(3\right),\left(4\right)\right)$  & $\left[2,2,2,3\right]$  & $\left(1,2,3,4\right)$\tabularnewline
\hline 
$\left[12,2\right]$  & $\left(\left(1\right),\left(2\right),\left(3,4\right)\right)$  & $\left[2,2,3,2\right]$  & $\left(1,2,3,4\right)$\tabularnewline
\hline 
\end{tabular}
\par\end{centering}

\caption{$\sigma_{1}$ and $\sigma_{2}$ values satisfying Eqn. \ref{eq:Coarse_Graining_Cycle}
for exchange symmetry in all bipartite decompositions of $N=24$.\label{tab:-bipartition_24}}
\end{table}

The cycle decomposition corresponding to cyclic shift of subsystems,
$\sigma_{c}=\left(1,2,\cdots,k\right)$ is related to that of bipartite
exchange symmetry by $\pi\left(\left[d_{1},d_{2},\cdots d_{k}\right],\sigma_{c}\right)=\pi\left(d^{'},d_{k}\right)$
where $d^{'}=\underset{i=1}{\overset{k-1}{\prod}}d_{i}$. Similarly,
$\pi\left(\left[d_{1},d_{2},\cdots d_{k}\right],\sigma_{c}^{-1}\right)=\pi\left(d_{1},d\right)$
where $d=\underset{i=2}{\overset{k}{\prod}}d_{i}$.

\subsection{\label{sub:Cyclic-invariance}Cyclic invariance in equi-dimensional
multipartitioning}

It may appear that the eigenvalues of $\hat{T}_{\mathbf{d},\sigma}$
not equal to $\pm1$ exist only when $\sigma\left(\mathbf{d}\right)\neq\mathbf{d}$,
that is, only when subsystems of distinct dimensions are permuted.
However, this is not the case. Consider an $k-$partite decomposition
$\mathbf{d}$ where all the subsystems are of equal dimensions $d$,
such that $N=d^{k}$. Given a TPS $\mathbb{C}^{\boldsymbol{d}}$,
consider the permutation $\sigma_{c}=\left(1,2,\cdots,k\right)$ which
is the cyclic permutation of $k-$subsystems where $k=n\left(\boldsymbol{d}\right)$:

\begin{equation}
\sigma_{c}\left(i\right)=\left(i+1\right)\mbox{ mod }k,\mbox{ for }i=1,\cdots,k.
\end{equation}

Given $k$ qudits, and $k$ parties $A_{1},A_{2},\cdots,A_{k}$, the
eigenstates of $\hat{T}_{\mathbf{d},\sigma_{c}}$ are such that their
interpretation remains identical irrespective of which qudit each
party makes the measurement on, as long as the measurements are done
in the order $A_{1},A_{2},\cdots,A_{k}$. Now, since $\hat{T}_{\mathbf{d},\sigma_{c}}$
and $\hat{T}_{\mathbf{d},\sigma_{c}^{-1}}$ share same eigenvectors,
these states have identical interpretation when the measurements are
carried out even in the anticyclic order $A_{1},A_{k},A_{k-1}\cdots,A_{2}$.
For example, consider $d=2$ and $k=4$, so that $\mathbf{d}=\left[2,2,2,2\right]$
and $\sigma_{c}=\left(1,2,3,4\right)$. The cycle decomposition $\pi\left(\left[2,2,2,2\right],\left(1,2,3,4\right)\right)$
is 
\[
\left(\left(0\right),\left(1,8,4,2\right),\left(3,9,12,6\right),\left(5,10\right),\left(7,11,13,14\right),\left(15\right)\right),
\]
from which the cyclic shift invariant states can be obtained. For
example, the $4-$cycle $\left(1,8,4,2\right)$ contributes $4$ eigenstates:
a symmetric state

\[
\frac{1}{2}\left(\left|0001\right\rangle +\left|1000\right\rangle +\left|0100\right\rangle +\left|0010\right\rangle \right),
\]

an anti-symmetric state

\[
\frac{1}{2}\left(-\left|0001\right\rangle +\left|1000\right\rangle -\left|0100\right\rangle +\left|0010\right\rangle \right),
\]

an eigenstate with eigenvalue $i$ :

\[
\frac{1}{2}\left(-i\left|0001\right\rangle -\left|1000\right\rangle +i\left|0100\right\rangle +\left|0010\right\rangle \right),
\]

and an eigenstate with eigenvalue $-i$:

\[
\frac{1}{2}\left(i\left|0001\right\rangle -\left|1000\right\rangle -i\left|0100\right\rangle +\left|0010\right\rangle \right).
\]

Symmetric subspace $\mathbb{S}_{\mathbf{d},\sigma}^{1}$ is six-dimensional
and the anti-symmetric subspace $\mathbb{S}_{\mathbf{d},\sigma}^{-1}$
is four-dimensional. The other two eigenspaces $\mathbb{S}_{\mathbf{d},\sigma}^{i}$
and $\mathbb{S}_{\mathbf{d},\sigma}^{-i}$ are both three-dimensional.

The eigenvalues of the cyclic shift operator and dimensions for the
corresponding eigenspaces for few $\mathbf{d}$ are shown in Table
\ref{tab:Cyclic_Shift_Examples} for illustration.

\begin{table}[H]
\begin{centering}
\begin{tabular}{|c|c|c|c|c|c|}
\hline 
$d$  & $k$  & $\mathbf{d}$  & $\sigma_{c}$  & Eigenvalues $e$  & Dimension of $\mathbb{S}_{\mathbf{d},\sigma_{c}}^{e}$\tabularnewline
\hline 
\hline 
$2$  & $3$  & $\left[2,2,2\right]$  & $\left(1,2,3\right)$  & $1,e^{\frac{2\pi i}{3}},e^{\frac{4\pi i}{3}}$  & $4,2,2$\tabularnewline
\hline 
$2$  & $4$  & $\left[2,2,2,2\right]$  & $\left(1,2,3,4\right)$  & $1,i,-1,-i$  & $6,3,4,3$\tabularnewline
\hline 
$3$  & $3$  & $\left[3,3,3\right]$  & $\left(1,2,3\right)$  & $1,e^{\frac{2\pi i}{3}},e^{\frac{4\pi i}{3}}$  & $11,8,8$\tabularnewline
\hline 
$3$  & $4$  & $\left[3,3,3,3\right]$  & $\left(1,2,3,4\right)$  & $1,i,-1,-i$  & $24,18,21,18$\tabularnewline
\hline 
$4$  & $3$  & $\left[4,4,4\right]$  & $\left(1,2,3\right)$  & $1,e^{\frac{2\pi i}{3}},e^{\frac{4\pi i}{3}}$  & $24,20$,20\tabularnewline
\hline 
$4$  & $4$  & $\left[4,4,4,4\right]$  & $\left(1,2,3,4\right)$  & $1,i,-1,-i$  & $70,60,66,60$\tabularnewline
\hline 
\end{tabular}
\par\end{centering}

\caption{Eigenvalues and dimensions of eigenspaces of circular permutation
invariant states of different $k$ and $d$.\label{tab:Cyclic_Shift_Examples}}
\end{table}

The eigenvalues of these permutations remain independent of $d$ and
depend only on $k$. Further, the cycle lengths in the cycle decomposition
$\pi\left(\mathbf{d},\sigma_{c}\right)$ are factors of $k$, so there
is no anti-symmetric subspace when $k$ is odd. It also follows that
if $k$ is prime then $\pi\left(\mathbf{d},\sigma_{c}\right)$ contains
$mod(d^{k}-2,k)+2$ number of $1-$cycles and $\left\lfloor \frac{d^{k}-2}{k}\right\rfloor $
number of $k-$cycles and no other cycles. Hence, the dimension of
the symmetric subspace in this case is $\left\lfloor \frac{d^{k}-2}{k}\right\rfloor +mod(d^{k}-2,k)+2$.

\section{\label{sec:Permutation-symmetry-and_Entanglement}Permutation symmetry
and Entanglement}

Entanglement of multipartite heterogenous states have been extensively
studied in the recent years. The standard notion of entanglement presupposes
an underlying TPS $\mathbb{C}^{\mathbf{d}}$. Given a TPS $\mathbb{C}^{\mathbf{d}}$,
a pure state $\left|\psi\right\rangle $ is separable if it is of
the form $\left|\psi_{1}\right\rangle \otimes\left|\psi_{2}\right\rangle \otimes\cdots\otimes\left|\psi_{k}\right\rangle $,
where $\left|\psi_{i}\right\rangle \in\mathbb{C}^{d_{i}}$. Otherwise,
the state is entangled. It is easy to see that entangled states in
a TPS need not be entangled in another. For instance, consider $\frac{1}{\sqrt{2}}\left(\left|1\right\rangle +\left|2\right\rangle \right)\in\mathbb{C}^{6}$.
Using the rule of association given in Section (??), this is identified
as $\left|0\right\rangle _{2}\otimes\frac{1}{\sqrt{2}}\left(\left|1\right\rangle _{3}+\left|2\right\rangle _{3}\right)\in\mathbb{C}^{2}\otimes\mathbb{C}^{3}$,
which is A poduct state. The corresponding state is $\frac{1}{\sqrt{2}}\left(\left|0\right\rangle _{3}\otimes\left|1\right\rangle _{2}+\left|1\right\rangle _{3}\otimes\left|0\right\rangle _{2}\right)\in\mathbb{C}^{3}\otimes\mathbb{C}^{2}$,
which is entangled.

As the focus of this work is on extending the notion of permutation
symmetry to heterogeneous systems, a suitable measure of entanglement
is required. Most of the multipartite entanglement measures exist
only in case of $d_{1}=d_{1}=\cdots=d_{k}=2$, that is, they are defined
only for $k-$partite qubit states. A recently proposed measure \cite{zhao2016new},
based on the degree of the mixedness of the reduced density matrices,
is

\begin{equation}
E_{t}\left(\left|\psi\right\rangle \right)=\underset{\left|A\right|=t}{\mbox{min}}\sqrt{\frac{d}{d-1}\left(1-\mbox{tr}\left(\rho_{A}^{2}\right)\right)}\;,d=\underset{i\in A}{\prod}d_{i}\label{eq:Entanglement_Formula}
\end{equation}
where $\left|\psi\right\rangle $ is an arbitrary $k$-qudit pure
state belonging to $\mathbb{C}^{d_{1}}\otimes\mathbb{C}^{d_{2}}\cdots\mathbb{C}^{d_{k}}$
and $t=1,2,\cdots,\left\lfloor \frac{k}{2}\right\rfloor $ where $\left\lfloor \frac{k}{2}\right\rfloor $
is the integral part of $\frac{k}{2}$ and $A$ is an arbitrary set
of $t$ qudits among the $k$ of them. Here $\rho_{A}=Tr_{\bar{A}}\left(\left|\psi\right\rangle \left\langle \psi\right|\right)$
is the reduced density matrix of the subsystem $A$. The quantity
$\sqrt{\frac{d}{d-1}\left(1-tr\left(\rho_{A}^{2}\right)\right)}$
measures the degree of mixedness associated with a specific bipartition
$\left\{ A|\overline{A}\right\} $ where $\overline{A}$ is the complement
of $A$. $E_{t}\left(\left|\psi\right\rangle \right)$ refers to the
minimum of this quantity among all possible bipartitions $\left\{ A|\overline{A}\right\} $
where $\left|A\right|=t$. For example, $E_{2}\left(\left|\psi\right\rangle \right)$
refers to the minimum of the entanglement existing every pair of systems
considered as a unit and the rest.

The maximally entangled state for a equi-dimensional $k-$partite
system is the generalized GHZ state,

\begin{equation}
\left|GHZ_{k,d}\right\rangle =\frac{1}{\sqrt{d}}\overset{d-1}{\underset{i=0}{\sum}}\left|\underset{k}{\underbrace{ii\cdots i}}\right\rangle _{\left[\underset{k}{\underbrace{d,d,\cdots,d}}\right]}=\frac{1}{\sqrt{d}}\overset{d-1}{\underset{i=0}{\sum}}\left|\alpha i\right\rangle ,\label{eq:max_ent_state}
\end{equation}
where $\alpha=\frac{d^{k}-1}{d-1}$. The prefactors of Eqn. This state
has an entanglement equal to $1$, with respect to the measure defined
in Eqn \ref{eq:Entanglement_Formula}. In the case of heterogeneous
$\mathbb{C}^{\mathbf{d}}$, a state of the form of Eqn. \ref{eq:max_ent_state},
with $d=min(\mathbf{d})$ is considered as a possible generalization.
Entanglement of this state is 
\begin{equation}
E_{1}\left(\left|GHZ_{k,d_{min}}\right\rangle \right)=\sqrt{\frac{d_{max}(d_{min}-1)}{d_{min}(d_{max}-1)}},\label{eq:Max_Ent_Formula}
\end{equation}
where $d_{min}=\mbox{min}(\mathbf{d})$ and $d_{max}=\mbox{max}(\mathbf{d})$.
This state is maximally entangled state when $k=2$, though the numerical
value of $E_{1}\left(\left|GHZ_{k,d_{min}}\right\rangle \right)$
measure is less than $1$. Further, the entanglement of this state
is identical in all decompositions $\sigma\left(\mathbf{d}\right)$,
for $\sigma\in S_{n\left(\mathbf{d}\right)}$.

\subsection{Bipartite exchange symmetry and entanglement}

\subsubsection{A measure of entanglement}

As $t=1$ for bipartite ($k=2$) decompositions, the entanglement
measure is denoted as $E$, without the subscript $t$. However, $E\left(\left|\psi\right\rangle \right)$
depends on the decomposition $\left[d_{1},d_{2}\right]$, which is
indicated with a suitable subscript as in $E\left(\left|\psi\right\rangle \right)$.
For example for the $\left[d_{1},d_{2}\right]$ bipartition,

\begin{equation}
_{\left[d_{1},d_{2}\right]}E\left(\left|\psi\right\rangle \right)=\sqrt{\frac{d_{max}}{d_{max}-1}\left(1-\mbox{Tr}\left(\rho_{i}^{2}\right)\right)},
\end{equation}
where $d_{min}=\mbox{max}(d_{1},d_{2})$ and $\rho_{i}$ could be
either of the reduced density matrices with $\left|\psi\right\rangle $
expressed in $\left[d_{1},d_{2}\right]$ partition. Similarly, $_{\left[d_{1},d_{2}\right]}E\left(\left|\psi\right\rangle \right)$
can be calculated. The entanglements differ in the way the reduced
density matrices are computed. The reduced density matrix of the first
subsystem after tracing over the second subsystem from $\mathbb{C}^{d_{1}}\otimes\mathbb{C}^{d_{2}}$
is: 
\begin{equation}
_{\left[d_{1},d_{2}\right]}\rho_{1}=\overset{d_{2}-1}{\underset{j=0}{\sum}}\left(\mathbb{I}_{d_{1}}\otimes\left\langle j\right|\right)\left|\psi\right\rangle \left\langle \psi\right|\left(\mathbb{I}_{d_{1}}\otimes\left|j\right\rangle \right)\label{eq:Rho1d1d2}
\end{equation}
where $\left\{ \left|j\right\rangle \right\} _{j=0}^{d_{2}-1}$ is
a basis for $\mathbb{C}^{d_{2}}$ and $\mathbb{I}_{d_{1}}$ is the
identity matrix in $\mathbb{C}^{d_{1}}$. In $_{\left[d_{1},d_{2}\right]}\rho_{1}$
notation, the prefix indicates the tensor product space and the suffix
indicates the subsystem in the factorization. The three other relevant
reduced density matrices are

\begin{equation}
_{\left[d_{2},d_{1}\right]}\rho_{2}=\overset{d_{2}-1}{\underset{j=0}{\sum}}\left(\left\langle j\right|\otimes\mathbb{I}_{d_{1}}\right)\left|\psi\right\rangle \left\langle \psi\right|\left(\left|j\right\rangle \otimes\mathbb{I}_{d_{1}}\right),\label{eq:Rho2d2d1}
\end{equation}

\begin{equation}
_{\left[d_{1},d_{2}\right]}\rho_{2}=\overset{d_{1}-1}{\underset{i=0}{\sum}}\left(\left\langle i\right|\otimes\mathbb{I}_{d_{2}}\right)\left|\psi\right\rangle \left\langle \psi\right|\left(\left|i\right\rangle \otimes\mathbb{I}_{d_{2}}\right),\label{eq:Rho2d1d2}
\end{equation}

\begin{equation}
_{\left[d_{2},d_{1}\right]}\rho_{1}=\overset{d_{1}-1}{\underset{i=0}{\sum}}\left(\mathbb{I}_{d_{2}}\otimes\left\langle i\right|\right)\left|\psi\right\rangle \left\langle \psi\right|\left(\mathbb{I}_{d_{2}}\otimes\left|i\right\rangle \right).\label{eq:Rho1d2d1}
\end{equation}

Of these four reduced density matrices, $_{\left[d_{1},d_{2}\right]}\rho_{1}$
and $_{\left[d_{2},d_{1}\right]}\rho_{2}$ are $d_{1}-$dimensional
whereas $_{\left[d_{2},d_{1}\right]}\rho_{2}$ and $_{\left[d_{1},d_{2}\right]}\rho_{2}$
are $d_{2}-$dimensional. For a generic $\left|\psi\right\rangle $,
$_{\left[d_{1},d_{2}\right]}\rho_{1}$ need not be equal to $_{\left[d_{2},d_{1}\right]}\rho_{2}$
and $_{\left[d_{1},d_{2}\right]}\rho_{2}$ need not be equal to $_{\left[d_{2},d_{1}\right]}\rho_{1}$.
Therefore, entanglement of these states, namely, $_{_{\left[d_{1},d_{2}\right]}}E\left(\left|\psi\right\rangle \right)$
and $_{\left[d_{2},d_{1}\right]}E\left(\left|\psi\right\rangle \right)$
are different. Nevertheless, if the state is exchange invariant, it
follows that 
\begin{equation}
_{\left[d_{1},d_{2}\right]}E\left(\left|\psi\right\rangle \right)=_{\left[d_{2},d_{1}\right]}E\left(\hat{T}_{\left[d_{1},d_{2}\right]}\left|\psi\right\rangle \right).\label{eq:Ent_D1_D2}
\end{equation}
One consequence of Eqn. \ref{eq:Ent_D1_D2} when $d_{1}=d_{2}=d$
is that the states $\left|\psi\right\rangle $ and $\hat{T}_{\left[d,d\right]}\left|\psi\right\rangle $
are equally entangled, for arbitrary $\left|\psi\right\rangle $.
Further, when $d_{1}\neq d_{2}$, the eigenstates of $\hat{T}_{\left[d_{1},d_{2}\right]}$
are equally entangled in both the partitions. This is a special case
of more general result. If $\left|\psi\right\rangle $ and $\hat{T}_{\left[d_{1},d_{2}\right]}\left|\psi\right\rangle $
are related as 
\begin{equation}
\hat{T}_{\left[d_{1},d_{2}\right]}\left|\psi\right\rangle =\hat{U}_{d_{2}}\otimes\hat{U}_{d_{1}}\left|\psi\right\rangle ,\label{eq:T_LU_Connection}
\end{equation}
where $\hat{U}_{d_{i}}$ is a local unitary operator of dimension
$d_{i}$, Eqn. \ref{eq:Ent_D1_D2} yields

\begin{equation}
_{\left[d_{1},d_{2}\right]}E\left(\left|\psi\right\rangle \right)=_{\left[d_{2},d_{1}\right]}E\left(\hat{U}_{d_{2}}\otimes\hat{U}_{d_{1}}\left|\psi\right\rangle \right)=_{\left[d_{2},d_{1}\right]}E\left(\left|\psi\right\rangle \right).
\end{equation}

not all the computational basis vectors are eigenstates of $\hat{T}_{\left[d_{1},d_{2}\right]}$
However they satisfy. \ref{eq:T_LU_Connection} and therefore, they
have equal entanglement $\left(=0\right)$ in both $\left[d_{1},d_{2}\right]$
and $\left[d_{2},d_{1}\right]$ bipartitions. It may be remarked that
the eigenstates of $\hat{T}_{\left[d_{1},d_{2}\right]}$ satisfy Eqn.
and are equally entangled in both the partitions. Thus, being an eigenstate
of the operator $\hat{T}_{\left[d_{1},d_{2}\right]}$ is sufficient
but not necessary for equally entangled in both the partitions.

Given a partition $\mathbb{C}^{d_{1}}\otimes\mathbb{C}^{d_{2}}$,
a basis set is defined as being of type $\left(p,q\right)$ if $p$
of the vectors are entangled and the rest $q=N-p$ basis vectors are
product states \cite{eakins2002factorization}. We could examine the
type of the privileged basis $\mathbb{B}_{\left[d_{1},d_{2}\right]}^{T}$
defined earlier. Since the elements of $\mathbb{B}_{\left[d_{1},d_{2}\right]}^{T}$
are equally entangled in both the partitions, its type would be same
in $\mathbb{C}^{d_{1}}\otimes\mathbb{C}^{d_{2}}$and $\mathbb{C}^{d_{2}}\otimes\mathbb{C}^{d_{1}}$
bipartitions. In the special case of qubit-qudit composite system,
it can be seen that $\mathbb{B}_{\left[2,d\right]}^{T}$ is always
of the type $\left(2d-2,2\right)$. Further, $\mathbb{B}_{\left[d,d\right]}^{T}$
is of $\left(d^{2}-d,d\right)$ type.

\subsubsection{Entanglement in the symmetric subspace}

The entanglement of the state $\left|\Gamma_{N}\right\rangle $, defined
in Eqn. \ref{eq:Gamma_State}, in $\left[d_{1},d_{2}\right]$ partition
is 
\begin{equation}
_{\left[d_{1},d_{2}\right]}E\left(\left|\Gamma_{N}\right\rangle \right)=\sqrt{\frac{d}{d-1}\frac{4\left(d_{1}-1\right)\left(d_{2}-1\right)-2}{\left(d_{1}d_{2}-2\right)^{2}}}\neq0\label{eq:E_Gamma_N}
\end{equation}
where $d=max(d_{1},d_{2})$. Therefore, $\left|\Gamma_{N}\right\rangle $
is entangled in every bipartition. For example, $\left|\Gamma_{4}\right\rangle $
is one of the Bell states, $\frac{1}{\sqrt{2}}\left(\left|01\right\rangle +\left|10\right\rangle \right)_{\left[2,2\right]}$,
which is maximally entangled in $\mathbb{C}^{2}\otimes\mbox{\ensuremath{\mathbb{C}}}^{2}$.

\subsubsection*{Product states in the symmetric subspace:}

Product states completely residing in the symmetric subspace of multipartite
qubit states are extensively studied in various contexts such as the
geometric measure of entanglement \cite{PhysRevA.68.042307}, qubit
spin coherent states in Majorana representation \cite{aulbach2010geometric},
etc. Here conditions on product state in $\mathbb{C}^{d_{1}}\otimes\mathbb{C}^{d_{2}}$
to belong to the symmetric subspace of $\hat{T}_{\left[d_{1},d_{2}\right]}$
are derived.

It is easy to see that product states $\left|0\right\rangle $ and
$\left|N-1\right\rangle $ belong to the symmetric subspace for every
bipartition of $N$. Consider the uniform state $\left|\Sigma_{N}\right\rangle $,
defined as

\begin{equation}
\left|\Sigma_{N}\right\rangle =\frac{1}{\sqrt{N}}\left(\stackrel[n=0]{N-1}{\sum}\left|n\right\rangle \right),\label{eq:Sigma_N}
\end{equation}
where $\left\{ \left|n\right\rangle \right\} _{n=0}^{N-1}$ is the
computational basis for $\mathbb{C}^{N}$ \cite{Wallach2008}. This
state differs from $\left|\Gamma_{N}\right\rangle $ defined in Eqn.
\ref{eq:Gamma_State}, in that the summation in $\left|\Sigma_{N}\right\rangle $
includes $\left|0\right\rangle $ and $\left|N-1\right\rangle $ also.
This state also belongs to the symmetric subspace (as it is a superposition
of symmetric states $\left|0\right\rangle $, $\left|\Gamma_{N}\right\rangle $
and $\left|N-1\right\rangle $), and is a product state in any bipartition
$\left[d_{1},d_{2}\right]$ as 
\begin{equation}
\left|\Sigma_{N}\right\rangle =\left(\frac{1}{\sqrt{d_{1}}}\stackrel[i=0]{d_{1}-1}{\sum}\left|i\right\rangle \right)\otimes\left(\frac{1}{\sqrt{d_{2}}}\stackrel[j=0]{d_{2}-1}{\sum}\left|j\right\rangle \right),
\end{equation}
where $\left\{ \left|i\right\rangle \right\} _{i=0}^{d_{1}-1}$ and
$\left\{ \left|j\right\rangle \right\} _{j=0}^{d_{2}-1}$ are the
computational bases of dimensions $d_{1}$ and $d_{2}$ respectively.
Hence states $\left|\Sigma_{N}\right\rangle ,\left|0\right\rangle $
and $\left|N-1\right\rangle $ are symmetric product states in every
partition. These product states in the symmetric subspace are refered
as trivial product states. It would be interesting to see whether
there are other product states in the symmetric subspace apart from
these trivial ones. That is, states $\left|\phi\right\rangle \in\mathbb{C}^{d_{1}}$
and $\left|\psi\right\rangle \in\mathbb{C}^{d_{2}}$ satisfying:

\begin{equation}
\left|\phi\right\rangle \otimes\left|\psi\right\rangle =\left|\psi\right\rangle \otimes\left|\phi\right\rangle .\label{eq:Symmetric_Prod_States}
\end{equation}
In case of $d_{1}=d_{2}=d$, symmetric product states are of the form

\begin{equation}
\left|\psi_{sep}^{sym}\right\rangle =\left|\epsilon\right\rangle \otimes\left|\epsilon\right\rangle ,
\end{equation}
where $\left|\epsilon\right\rangle \in\mathbb{C}^{d}$. When $d_{1}\neq d_{2}$,
finding states satisfying Eqn. \ref{eq:Symmetric_Prod_States} is
more involved \cite{horn1994topics}. Cycle decomposition will aid
in identifying the symmetric product states.

An arbitrary product state in the $\left[d_{1},d_{2}\right]$ bipartition
can be written in the computation basis as:

\begin{equation}
\left|\phi\right\rangle =\left(\overset{d_{1}-1}{\underset{i=0}{\sum}}\alpha_{i}\left|i\right\rangle \right)\otimes\left(\overset{d_{2}-1}{\underset{j=0}{\sum}}\beta_{j}\left|j\right\rangle \right)=\underset{i,j}{\sum}\alpha_{i}\beta_{j}\left|ij\right\rangle _{\left[d_{1},d_{2}\right]}\label{eq:Product_State}
\end{equation}
where $\alpha_{i}$ and $\beta_{j}$ are complex numbers, such that
$\overset{d_{1}-1}{\underset{i=0}{\sum}}\left|\alpha_{i}\right|^{2}=1$
and $\overset{d_{2}-1}{\underset{i=0}{\sum}}\left|\beta_{i}\right|^{2}=1$.
For $\left|\phi\right\rangle $ to be an eigenstate of $T_{\left[d_{1},d_{2}\right]}$,
$\alpha_{i}$ and $\beta_{j}$ need to satisfy constraints arising
due to each cycle in $\pi\left(d_{1},d_{2}\right)$. Consider one
of the cycles $\left(L_{1},L_{2},\cdots,L_{l}\right)$ in $\pi\left(d_{1},d_{2}\right)$.
Recall that $L_{1},L_{2}..,L_{l}$ are all integers between $0$ and
$d_{1}d_{2}-1$. For notational convenience, we use the following
symbols $\widehat{x}\equiv\left\lfloor \frac{x}{d_{2}}\right\rfloor $
and $\overline{x}\equiv\mbox{mod}\left(x,d_{2}\right)$ so that the
state $\left|L_{r}\right\rangle $ in $\left[d_{1},d_{2}\right]$
decomposition is $\left|\widehat{L_{r}},\overline{L_{r}}\right\rangle _{\left[d_{1},d_{2}\right]}$,
and from Eqn. \ref{eq:Product_State} it can be seen that in the expansion
of $\left|\phi\right\rangle $, the coefficient of $\left|\widehat{L_{r}},\overline{L_{r}}\right\rangle _{\left[d_{1},d_{2}\right]}$
is $\alpha_{\widehat{L_{r}}}\beta_{\overline{L_{r}}}$. For the state
$\left|\phi\right\rangle $ to remain invariant under $\hat{T}_{\left[d_{1},d_{2}\right]}$,
the complex coefficients $\alpha_{i}$ and $\beta_{j}$ of Eqn. \ref{eq:Product_State}
have to satisfy the following constraints:

\begin{equation}
\alpha_{\widehat{L_{1}}}\beta_{\overline{L_{1}}}=\alpha_{\widehat{L_{2}}}\beta_{\overline{L_{2}}}=\cdots=\alpha_{\widehat{L_{l}}}\beta_{\overline{L_{l}}}.\label{eq:Product_States_Constraints}
\end{equation}

For every cycle of length $l$ greater than $1$, there are $\left(\begin{array}{c}
l\\
2
\end{array}\right)$ similar such equalities on the coefficients $\alpha_{i}$ and $\beta_{j}$.
For example, consider the $\left[2,3\right]$ partition which has
$\pi\left(2,3\right)=\left(\left(0\right),\left(1,2,4,3\right),\left(5\right)\right)$.
Consider one of the cycles of $\pi\left(2,3\right)$, say $\left(1,2,4,3\right)$.
For product state$\left|\phi\right\rangle $ to be a symmetric state,
the coefficients $\alpha_{i}$ and $\beta_{j}$ are required to satisfy
(see Eqn. \ref{eq:Product_States_Constraints}) the following three
independent constraints:

\begin{equation}
\alpha_{0}\beta_{1}=\alpha_{0}\beta_{2}=\alpha_{1}\beta_{1}=\alpha_{1}\beta_{0}.\label{eq:Symm_Prod_Constrsaint_2_3}
\end{equation}
The other two cycles $\left(0\right)$ and $\left(5\right)$ correspond
to symmetric eigenstates by themselves and do not yeild any additional
constraints. The only state satisfying these three constraints is
$\left|\Sigma_{6}\right\rangle $. There are no other symmetric product
states in $\mathbb{C}^{2}\otimes\mathbb{C}^{3}$ apart from $\left|0\right\rangle ,\left|5\right\rangle $
and $\left|\Sigma_{6}\right\rangle $. In fact, for situations where
$\pi\left(d_{1},d_{2}\right)$ has only three cycles (that is two
$1-$cycles and one $d_{1}d_{2}-2$ cycle; see for example $\pi\left(2,6\right)$
in Table \ref{tab:-Pi_Examples}), it is easy to see that there are
no other symmetric product states apart from the trivial ones.

On the other hand, $\mathbb{C}^{2}\otimes\mathbb{C}^{4}$ has $\pi\left(2,4\right)=\left(\left(0\right),\left(1,2,4\right),\left(3,6,5\right),\left(7\right)\right)$.
The cycle $\left(1,2,4\right)$ offers three constraints $\alpha_{0}\beta_{1}=\alpha_{0}\beta_{2}=\alpha_{1}\beta_{0}$
and the cycle $\left(3,6,5\right)$ contributes three more constraints,
$\alpha_{0}\beta_{3}=\alpha_{1}\beta_{1}=\alpha_{1}\beta_{2}$. These
six constraints are satisfied provided $\beta_{1}=\beta_{2}$, $\alpha_{0}\beta_{1}=\alpha_{1}\beta_{0}$
and $\alpha_{0}\beta_{3}=\alpha_{1}\beta_{1}$.

States appearing as fixed-points are symmetric product states. For
example, $\pi\left(3,5\right)$ (see Table \ref{tab:-Pi_Examples})
has state $\left|7\right\rangle $ appearing as a $1-$cycle, which
in $\mathbb{C}^{3}\otimes\mathbb{C}^{5}$ decomposition is $\left|1\right\rangle _{3}\otimes\left|2\right\rangle _{5}$
and in $\mathbb{C}^{5}\otimes\mathbb{C}^{3}$ decomposition is $\left|2\right\rangle _{5}\otimes\left|1\right\rangle _{3}$.
Incidentally, $\mathbb{C}^{3}\otimes\mathbb{C}^{5}$ is the smallest
(in terms of $N$) heterogenous bipartitite TPS where $\pi\left(d_{1},d_{2}\right)$
has $1-$cycles other than $\left|0\right\rangle $ and $\left|d_{1}d_{2}-1\right\rangle $:
in other words smallest $d_{1}$ and $d_{2}$ $\left(\neq d_{1}\right)$
for which the matrix $\hat{T}_{\left[d_{1},d_{2}\right]}$ has trace
greater than two. It follows from Eqn. \ref{eq:N_l} that when $d_{1}$
or $d_{2}$ is 2, $\sigma\left(1\right)$ is $2$. In that case, there
are only two fixed points in $\pi\left(2,d\right)$ and $\pi\left(d,2\right)$.
Similarly, cycle decomposition $\pi\left(3,4\right)$ also has no
cycle of length one apart from $\left(0\right)$ and $\left(11\right)$,
see Table \ref{tab:-Pi_Examples}.

When $N$ is of the form $d^{k}$, where $d$ is a prime number, then
recall that the symmetric product states in the homogenous $k-$partite
decomposition are of the form $\left|\epsilon\right\rangle \otimes\left|\epsilon\right\rangle \otimes\cdots\left|\epsilon\right\rangle $,
where $\left|\epsilon\right\rangle \in\mathbb{C}^{d}$ is a normalized
pure state. Now, it is easy to see these states would remain symmetric
product states in any coarse grained decomposition $\mathbb{C}^{\mathbf{d}}$,
where $\mathbf{d}\in\mathbb{P}\left(N\right)$.

\subsubsection{Entanglement in the non-symmetric eigenspaces of $\hat{T}_{\left[d_{1},d_{2}\right]}$}

The central result of this paper is the observation that the non-symmetric
eigenspaces of $\hat{T}_{\left[d_{1},d_{2}\right]}$, $\mathbb{S}_{\Xi,\sigma}^{\eta},\:\eta\neq1$
are completely entangled. There are no product states in either partitioning
in these subspaces. To see this, assume on the contrary that a $\left[d_{1},d_{2}\right]$
product state $\left|\psi\right\rangle \otimes\left|\phi\right\rangle $
belongs to the non-symmetric eigenspace of $\hat{T}_{\left[d_{1},d_{2}\right]}$.
Then $\hat{T}_{\left[d_{1},d_{2}\right]}\left(\left|\psi\right\rangle \otimes\left|\phi\right\rangle \right)=e^{\frac{2\pi ik}{n}}\left|\psi\right\rangle \otimes\left|\phi\right\rangle $,
for some integers $n$ and $k$ such that $0<k<n$. But this is impossible
as the real matrix $\hat{T}_{\left[d_{1},d_{2}\right]}$ only permutes
the entries of $\left|\psi\right\rangle \otimes\left|\phi\right\rangle $
and cannot introduce a complex phase. It is known that non-symmetric
eigenspaces of $T_{\left[d,d\right]}$ are completely entangled \cite{PhysRevA.78.052105}.
Our result generalization to heterogenous systems.

As eigenstates of $\hat{T}_{\left[d_{1},d_{2}\right]}$ have equal
entanglement in both $\mathbb{C}^{d_{1}}\otimes\mathbb{C}^{d_{2}}$
and $\mathbb{C}^{d_{2}}\otimes\mathbb{C}^{d_{1}}$, the non-symmetric
eigenspaces of $\hat{T}_{\left[d_{1},d_{2}\right]}$ are completely
entangled subspaces in both of them. This way, given $d_{1}$ and
$d_{2}$, one obtains as many completely entangled subspaces as there
are distinct non-unit eigenvalues of $\hat{T}_{\left[d_{1},d_{2}\right]}$,
given by Eqn. \ref{eq:Eigen_Spectrum}.

The largest subspace of a TPS where every vector is entangled is discussed
in \cite{parthasarathy2004maximal} and an explicit construction of
a basis for such a subspace is provided in \cite{bhat2006completely}.
Given $d_{1}$ and $d_{2}$, the largest completely entangled subspaces
(CES) in $\mathbb{C}^{d_{1}}\otimes\mathbb{C}^{d_{2}}$ and $\mathbb{C}^{d_{2}}\otimes\mathbb{C}^{d_{1}}$
are $\left(d_{1}-1\right)\left(d_{2}-1\right)$ dimensional\cite{parthasarathy2004maximal}.

Given $\mathbb{C}^{d_{1}}\otimes\mathbb{C}^{d_{2}}$, the largest
CES is the one orthogonal to $R^{\perp}$ given as \cite{bhat2006completely}:
\begin{equation}
R^{\perp}=\mbox{span}\left\{ P\underset{i_{1}+i_{2}=n}{\sum}\left|i_{1}\right\rangle \otimes\left|i_{2}\right\rangle \mbox{,}n=0,\cdots,n_{max}\right\} ,\label{eq:SBar}
\end{equation}
where $\left\{ \left|i_{1}\right\rangle \right\} _{i_{1}=0}^{d_{1}-1}$
is an orthonormal basis in $\mathbb{C}^{d_{1}}$, $\left\{ \left|i_{2}\right\rangle \right\} _{i_{2}=0}^{d_{2}-1}$
is an orthonormal basis in $\mathbb{C}^{d_{2}}$, $n_{max}=d_{1}+d_{2}-2$
and $P$ is the normalization constant. 

The largest CES $R_{\left[d_{1},d_{2}\right]}$ and $R_{\left[d_{2},d_{1}\right]}$
are related as 
\begin{equation}
R_{\left[d_{2},d_{1}\right]}=\hat{T}_{\left[d_{1},d_{2}\right]}R_{\left[d_{1},d_{2}\right]},\label{eq:R_Conversion}
\end{equation}
where $\hat{T}_{\left[d_{1},d_{2}\right]}R_{\left[d_{1},d_{2}\right]}$
stands for the subspace spanned by the vectors of the form $\hat{T}_{\left[d_{1},d_{2}\right]}\left|\psi\right\rangle $
where $\left\{ \left|\psi\right\rangle \right\} $ span $R_{\left[d_{1},d_{2}\right]}.$
A subscript is used to $R$ to denote the TPS in which it is completely
entangled. Note that vectors in $R_{\left[d_{1},d_{2}\right]}$ need
not be entangled when viewed as states in $\left[d_{2},d_{1}\right]$
partition and vice-versa.

Given two CES $R_{\left[d_{1},d_{2}\right]}$ and $R_{\left[d_{2},d_{1}\right]}$,
their intersection $R_{\left[d_{1},d_{2}\right]}\cap R_{\left[d_{2},d_{1}\right]}$
is also a CES in which every vector is entangled in both $\mathbb{C}^{d_{1}}\otimes\mathbb{C}^{d_{2}}$
and $\mathbb{C}^{d_{2}}\otimes\mathbb{C}^{d_{1}}$. The states in
the intersection, however, generally have different entanglement in
the two TPSs. The non-symmetric eigenspaces of $T_{\left[d_{1},d_{2}\right]}$,
on the other hand, are CES in which every vector is equally entangled
in both the partitions.

Again, to make progress we study qubit-qudit bipartite TPS. The largest
CES subspaces in $[2,d]$ and $[d,2]$ partitions are both $d-1$
dimensional, given by

\begin{eqnarray}
R_{\left[2,d\right]} & = & \mbox{span}\left\{ \frac{1}{\sqrt{2}}\left(\left|0.i\right\rangle -\left|1.(i-1)\right\rangle \right)_{\left[2,d\right]}\right\} ,\label{eq:CES_2d}\\
R_{\left[d,2\right]} & = & \mbox{span}\left\{ \frac{1}{\sqrt{2}}\left(\left|(i-1).1\right\rangle -\left|i.0\right\rangle \right)_{\left[d,2\right]}\right\} ,\label{eq:CES_d2}
\end{eqnarray}
where $i$ runs from $1$ to $d-1$.

The basis vectors of $R_{\left[2,d\right]}\left(\mbox{resp }R_{\left[d,2\right]}\right)$
are all equally entangled in the $\left[2,d\right]\left(\mbox{resp }\left[d,2\right]\right)$
partition with $E=\sqrt{\frac{d}{2d-2}}$, which is the maximum entanglement
in the $\left[2,d\right]\left(\mbox{resp }\left[d,2\right]\right)$
partition (see Eqn. \ref{eq:Max_Ent_Formula}). The dimension of the
intersection of $R_{\left[2,d\right]}$ and $R_{\left[d,2\right]}$
subspaces depends on whether $d$ is odd or even. If $d$ is odd,
the intersection is $\frac{d-1}{2}$ dimensional, and it is the span
of $\left\{ \frac{1}{2}\left(\left|2i-1\right\rangle -\left|2i\right\rangle -\left|d+2i-2\right\rangle +\left|d+2i-1\right\rangle \right)\right\} \mbox{, for }i=1,2,...,\frac{d-1}{2}$.
If $d$ is even, it is one-dimensional, spanned by

\begin{equation}
R_{[2,d]}\cap R_{[d,2]}=\frac{1}{\sqrt{2(d-1)}}\left(\stackrel[i=1]{2(d-1)}{\sum}\left(-1\right)^{i+1}\left|i\right\rangle \right),\mbox{even }d\label{eq:R2d_d2_Intersection_1D}
\end{equation}

When $d_{1}=d_{2}=d$, there is only one non-symmetric eigenspace,
the $\frac{1}{2}d\left(d-1\right)$ dimensional anti-symmetric subspace
$A_{\left[d,d\right]}$ given by:

\begin{equation}
A_{\left[d,d\right]}=\mbox{span}\left\{ \frac{1}{\sqrt{2}}\left(\left|i\right\rangle \otimes\left|j\right\rangle -\left|j\right\rangle \otimes\left|i\right\rangle \right)\right\} ,\label{eq:A_Symm_Bipart}
\end{equation}
for $i,j\in(0,\cdots,d-1)$ and $i>j$. In this case, $A_{\left[d,d\right]}\subseteq R_{\left[d,d\right]}$
with the equality holding only when $d=2$. 

All the basis vectors of $A_{\left[d,d\right]}$ listed above have
$\mbox{Tr}\left(\rho_{A}^{2}\right)=\frac{1}{2}$, for all $d$. Hence,
entanglement of any of the basis vectors is $\sqrt{\frac{d}{2\left(d-1\right)}}$.
Further, it has been numerically verified (for over $10^{4}$ states,
sampled randomly with respect to Haar measure \cite{ozols2009generate})
that the lowest entanglement in the anti-symmetric subspace of $\mathbb{C}^{d}\otimes\mathbb{C}^{d}$
is $\sqrt{\frac{d}{2\left(d-1\right)}}$.

\subsection{Multipartite permutation symmetry and Entanglement}

For a general state $\left|\psi\right\rangle $, a decomposition $\mathbf{d}$
and a permutation $\sigma$, analogous to Eqn. \ref{eq:Ent_D1_D2},
the following relation holds:

\begin{equation}
_{\mathbf{d}}E\left(\left|\psi\right\rangle \right)=_{\sigma\left(\mathbf{d}\right)}E\left(\hat{T}_{\mathbf{d},\sigma}\left|\psi\right\rangle \right),\label{eq:Ent_D1_D2_MP}
\end{equation}
for all $t=1,2,\cdots,\left\lfloor \frac{k}{2}\right\rfloor $ in
Eqn. \ref{eq:Entanglement_Formula}. As in the bipartite case (Eqn.
\ref{eq:T_LU_Connection}), if $\left|\psi\right\rangle $ and $\hat{T}_{\mathbf{d},\sigma}\left|\psi\right\rangle $
are related as:

\begin{equation}
\hat{T}_{\mathbf{d},\sigma}\left|\psi\right\rangle =\underset{i}{\otimes}\hat{U}_{\sigma^{-1}\left(i\right)}\left|\psi\right\rangle ,
\end{equation}
where $\hat{U}_{r}$ is the local unitary transformation of dimension
$d_{r}$, then eqn. \ref{eq:Ent_D1_D2_MP} is satisfied.

For a given $N$, the states $\left|0\right\rangle $, $\left|N-1\right\rangle $,
$\left|\Gamma_{N}\right\rangle $ and $\left|\Sigma_{N}\right\rangle $
belong to the symmetric subspace in $\mathbb{C}^{\mathbf{d}}$, for
any $\mathbf{d}\in\mathbb{P}\left(N\right)$ and any $\sigma\in S_{n\left(\mathbf{d}\right)}$.
Of these, states $\left|0\right\rangle $, $\left|N-1\right\rangle $
and $\left|\Sigma_{N}\right\rangle $ are product states in every
partition $\mathbf{d}$, whereas $\left|\Gamma_{N}\right\rangle $
is entangled. The entanglement in the later is given by 
\begin{equation}
_{\mathbf{d}}E\left(\left|\Gamma_{N}\right\rangle \right)=\sqrt{\frac{d}{d-1}\frac{4\left(d-1\right)\left(d^{\prime}-1\right)-2}{\left(N-2\right)^{2}}},\label{eq:Ent_Gamma_N_Multipartite}
\end{equation}
where $d=max\left(\mathbf{d}\right)$ and $d^{\prime}=\frac{N}{d}$. 

\begin{figure}[h]
\begin{raggedright} %
\begin{minipage}[t]{0.45\textwidth}%
\begin{center}
$a)$\includegraphics[scale=0.3]{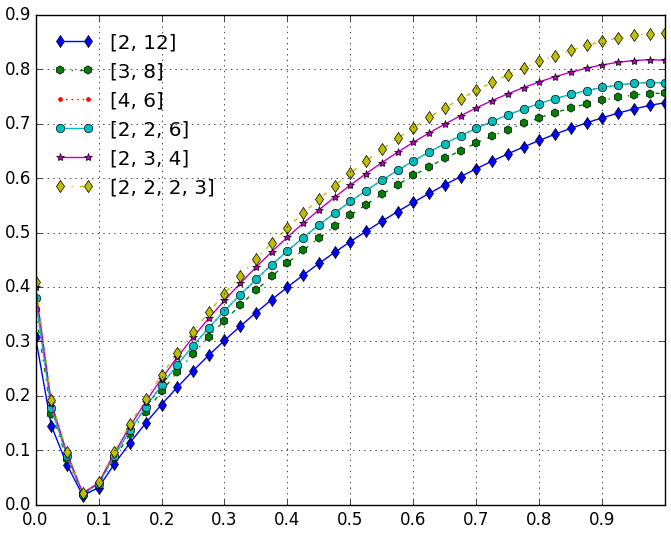} 
\par\end{center}%
\end{minipage}\hfill{}%
\begin{minipage}[t]{0.45\textwidth}%
\begin{center}
$b)$\includegraphics[scale=0.3]{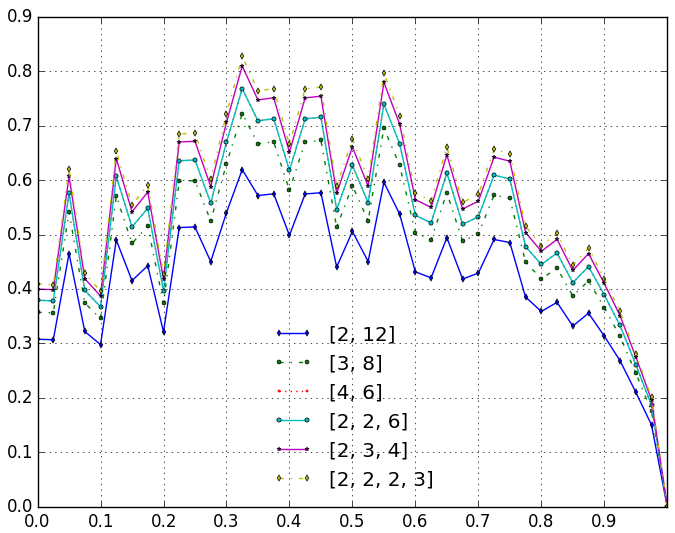} 
\par\end{center}%
\end{minipage}

\end{raggedright}

\caption{Entanglement of the states a) $\left|\chi_{1}\left(p\right)\right\rangle $
and b) $\left|\chi_{2}\left(p\right)\right\rangle $ of Eqn. \ref{eq:Chi_P_Al_Ent}
as function of $p$, in all representative decompositions \label{fig:Chi_P_All_Ent}
in $\mathbb{P}\left(24\right)$.}
\end{figure}

It will be instructive to examine entanglement of states in the generalized
symmetric subspace, defined in Eqn \ref{eq:GSS}, in all representative
partitions $\mathbf{d}_{e}$. Consider two families of states:

\begin{eqnarray}
\left|\chi_{1}\left(p\right)\right\rangle  & = & \sqrt{p}\frac{1}{\sqrt{2}}\left(\left|0\right\rangle +\left|N-1\right\rangle \right)+\sqrt{1-p}\left|\Gamma_{N}\right\rangle \nonumber \\
\left|\chi_{2}\left(p\right)\right\rangle  & = & \sqrt{p}\left|0\right\rangle +e^{i\phi\left(p\right)}\sqrt{1-p}\left|\Gamma_{N}\right\rangle \label{eq:Chi_P_Al_Ent}
\end{eqnarray}
where relative phase $\phi\left(p\right)$ is a random variable between
$0$ to $2\pi$. Figure \ref{fig:Chi_P_All_Ent} shows the variation
of entanglement of these two families of states for $0\leq p\leq1$
and $N=24$. These states belong to the symmetric subspace for all
permutations $\sigma$, therefore it enough to study their entanglement
in the representative decompositions of $N$.

There are six representative factorizations of $\mathbb{C}^{24}$:
three bipartite, two tripartite, and the four-partite primitive decomposition.
Variation of entanglement with $p$ of these families of states is
given in Fig \ref{fig:Chi_P_All_Ent}. Entanglement is largest in
the primitive decomposition $\mathbf{d}_{p}=\left[2,2,2,3\right]$
and least in the $\left[2,12\right]$ decomposition for both $\left|\chi_{1}\left(p\right)\right\rangle $
and $\left|\chi_{2}\left(p\right)\right\rangle $, for all values
of $p$.

For $p=0$,$\left|\chi_{1}\right\rangle $ is $\left|\Gamma_{N}\right\rangle $,
which is entangled in every partition of $N$ (see Eqn. \ref{eq:Ent_Gamma_N_Multipartite}).
At $p=1$, it corresponds to the GHZ-like state having equal superposition
of two product states $\left|0\right\rangle $ and $\left|N-1\right\rangle $.
From Fig. \ref{fig:Chi_P_All_Ent}(a) it is seen that this state is
more entangled than $\left|\Gamma_{N}\right\rangle $. At $p=\frac{2}{N}$,
the state $\left|\chi_{1}\right\rangle $ is $\left|\sum_{N}\right\rangle $
of Eqn. \ref{eq:Sigma_N}, which is a product state in every decomposition
$\mathbf{d}$, which explains the dip at $p=\frac{1}{12}$ for $N=24$
in all the plots of Fig. \ref{fig:Chi_P_All_Ent}(a).

Fig. \ref{fig:Chi_P_All_Ent}(b) is plot of entanglement in the states
$\left|\chi_{2}\left(p\right)\right\rangle $, which are superpositions
of the product state $\left|0\right\rangle $ and $\left|\Gamma_{N}\right\rangle $
with a random relative phase. It is evident that entanglement of $\left|\chi_{2}\left(p\right)\right\rangle $
shows identical variation with $p$ in all TPSs. These observations
are independent of $N$. Here $N=24$ was chosen only because it has
a number of ditinct partitions.

So far, entanglement in the symmetric subspace has been discussed.
Now, entanglement in the nonsymmetric eigenspaces $\mathbb{S}_{\boldsymbol{d},\sigma}^{\eta}$,
$\eta\neq1$ of $\hat{T}_{\mathbf{d},\sigma}$ will be examined. As
an illustration, consider $\boldsymbol{d}=\left[2,2,3\right]$ and
$\sigma=\left(\left(1,2\right),\left(3\right)\right)$. There are
three cycles of even lengths in the cycle decomposition $\pi\left(\left[2,2,3\right],\left(\left(1,2\right),\left(3\right)\right)\right)$
(see the first row of Table \ref{tab:Tripartite}). This implies that
the anti-symmetric subspace is three dimensional:

\begin{eqnarray*}
\mathbb{S}_{\boldsymbol{d},\sigma}^{-1} & = & \mbox{span}\left\{ \begin{array}{c}
\frac{1}{\sqrt{2}}\left(\left|010\right\rangle -\left|100\right\rangle \right)_{\left[2,2,3\right]}\\
\frac{1}{\sqrt{2}}\left(\left|011\right\rangle -\left|101\right\rangle \right)_{\left[2,2,3\right]}\\
\frac{1}{\sqrt{2}}\left(\left|012\right\rangle -\left|102\right\rangle \right)_{\left[2,2,3\right]}
\end{array}\right\} 
\end{eqnarray*}

Subspace $\mathbb{S}_{\boldsymbol{d},\sigma}^{-1}$ is a CES in the
sense that there are no product state of the form $\left|\alpha_{1}\right\rangle \otimes\left|\alpha_{2}\right\rangle \otimes\left|\beta\right\rangle $
in this subspace where $\left|\alpha_{i}\right\rangle \in\mathbb{C}^{2}$
and $\left|\beta\right\rangle \in\mathbb{C}^{3}$. But states in $\mathbb{S}_{\boldsymbol{d},\sigma}^{-1}$
are entangled only with respect to the first and second subsystems.
Therefore, there is no genuine tripartite entanglement in this subspace.
Indeed, the entanglement of states in this subspace with respect to
the measure Eqn. \ref{eq:Entanglement_Formula} is zero:

\begin{equation}
E_{1}\left(\left|\psi\right\rangle \right)=0\:\forall\left|\psi\right\rangle \in\mathbb{S}_{\boldsymbol{d},\sigma}^{-1}
\end{equation}

Similarly, there is no genuine tripartite entanglement in subspaces
$\mathbb{S}_{\boldsymbol{d},\sigma}^{-1}$ for $\sigma=\left(\left(1,3\right),\left(2\right)\right)$
and $\sigma=\left(\left(1\right),\left(2,3\right)\right)$. It can
be inferred from this example that if $\sigma$ involves permutation
of only a subset of subsystems, the corresponding non-symmetric eigenspaces
will be genuinely entangled only with respect to those subsystems.
The states in the subspace will be separable with respect to the rest
of the subsystems.

Now, consider a permutation $\sigma$ such that $\sigma\left(i\right)\neq i$
for $i=1,\cdots,k$. In this case, the non-symmetric eigenspaces of
$T_{\mathbf{d},\sigma}$ are all completely entangled in both $\mathbf{d}$
and $\sigma\left(\mathbf{d}\right)$ partitions. For example, $\pi\left(\left[2,2,3\right],\left(1,3,2\right)\right)$
(see last row of Table \ref{tab:Tripartite}) has one even length
cycle. The corresponding anti-symmetric state is genuienly entangled.
For this state, the quantum of entanglement with respect to the measure
defined in Eqn. \ref{eq:Entanglement_Formula} is $0.9$. Permutation
$\sigma_{c}$ discussed in section \ref{sub:Cyclic-invariance} is
another example where the nonsymmetric eigenspaces are genuinely multipartite
entangled. To the best of our knowledge, there is no other prescription
for generating genuinely completely entangled subspaces. For example,
the construction discussed in \cite{bhat2006completely}, in case
of $k$ partite qubit system, generates the subspace orthogonal to
the conventional symmetric subspace (the space spanned by the Dicke
basis). This CES is $2^{k}-(k+1)$ dimensional, but it has states
which do not have genuine entanglement.

\section{\label{sec:Summary}Summary}

Symmetry is one of the fundamental notions in physics, and its role
in quantum mechanics cannot be overstated. In multipartite quantum
systems, a natural symmetry operation is permutation symmetry. For
homogenous $k-$partite systems, one identifies the ``symmetric subspace''
as the span of the states that remain invariant under any permutation
of the subsystem labels. 

Permutation symmetry of multipartite quantum states is generally considered
only in the homogenous setting. A way of extending this symmetry to
the case when subsystems are of unequal dimensions has been established
here. This extension has been achieved via the natural isomorphism
existing between the unfactored Hilbert space and the tensor product
of the heterogeneous subsystems taken in different ordering. This
extension recovers the conventional definition of permutation symmetry
in the homogenous case. This has been accomplished by extending the
idea of permutation matrix in the bipartite homogeneous case to multipartite
heterogenous case. In the computational basis of $\mathbb{C}^{N}$,
these matrices are permutation matrices. An algorithm for obtaining
the permutations $\pi\in S_{N}$, corresponding to these matrices
has been provided. The eigenvectors of $\hat{T}_{\mathbf{d},\sigma}$
are such that they have identical representation in both the tensor
product spaces $\mathbb{C}^{\mathbf{d}}$ and $\mathbb{C}^{\sigma\left(\mathbf{d}\right)}$.
The eigenspaces of $\hat{T}_{\mathbf{d},\sigma}$ corresponding to
eigenvalue $+1$ are symmetric subspaces and eigenvalue $-1$ are
anti-symmetric subspaces. This definition is meaningful as it gives
rise to the conventional notions of symmetric and anti-symmetric states
when $\mathbf{d}=\sigma\left(\mathbf{d}\right)$, which is possible
if the system is homogeneous or the permutation is among the subsystems
of equal dimensions. Moreover, this extension gives rise to classes
of states other than the symmetric and antisymmetric ones. These are
states which acquire a global complex phase $(\neq\pm1)$ under action
of $\hat{T}_{\mathbf{d},\sigma}$. A procedure to obtain the dimension
of each of these eigenspaces of $\hat{T}_{\mathbf{d},\sigma}$ by
examining the corresponding permutation $\pi\left(\mathbf{d},\sigma\right)$
has been discussed. Further, it has been shown that all the nonsymmetric
eigenspaces (i.e., eigenspaces corresponding to eigenvalues $\neq1$)
of $\hat{T}_{\mathbf{d},\sigma}$ are completely entangled subspaces.
There are no product states in these subspaces. Further, these states
have equal entanglement in both $\mathbb{C}^{\mathbf{d}}$ and $\mathbb{C}^{\sigma\left(\mathbf{d}\right)}$.
These completely entangled subspaces are distinct from those discussed
by Bhat \cite{bhat2006completely}. If $\sigma$ is such that it has
no cycles of length one, the states in these completely entangled
subspaces are also genuinely entangled in the sense they remain entangled
under arbitrary bipartitions.

For a given unfactored space of dimension $N$, we have identified
a unique tensor product space composed of subspaces whose dimensions
are the prime factors of $N$, tensored in the order of increasing
subsystem dimensions. This unique tensor product space has the maximum
number of subsystems and every other coarse-grained tensor product
space consistent with $N$ can be obtained by permutation (if needed)
and merging of the subsystems of this unique factorzation. It has
been established that the permutation symmetries of such coarse-grained
tensor product spaces are expressible in terms of the permutation
symmetries of this unique tensor product space. 
\begin{acknowledgments}
We thank Ludovic Arnaud for his insightful feedback on the manuscript.
We also thank A.K. Rajgopal, Ajit Iqbal Singh and D. Goyeneche for
their useful comments. 
\end{acknowledgments}

\appendix
\begin{table*}[p]
\begin{centering}
\begin{tabular}{|c|>{\raggedright}m{15cm}|}
\hline 
\textbf{Symbol}  & \textbf{Description}\tabularnewline
\hline 
\hline 
$\mathbb{C}^{N}$  & Complex vector space of dimension $N$\tabularnewline
\hline 
$\left[d_{1},d_{2}\right]$  & A bipartite decomposition of $N=d_{1},d_{2}$. \tabularnewline
\hline 
$\left|i\right\rangle _{d_{1}}$  & $\left(i+1\right)^{th}$ computational basis vector in $\mathbb{C}^{d_{1}}$.
A $d_{1}-$dimensional column vector having $1$ in $\left(i+1\right)^{th}$
position and $0$ everywhere else. \tabularnewline
\hline 
$\mathbb{B}_{d_{j}}$  & Computational basis of $\mathbb{C}^{d_{j}}$.\tabularnewline
\hline 
$\mathbb{B}$  & Computational basis of $\mathbb{C}^{N}$.\tabularnewline
\hline 
$\hat{T}_{\left[d_{1},d_{2}\right]}$  & Subsystem permutation operator mapping product state $\left|a\right\rangle \otimes\left|b\right\rangle $
in $\mathbb{C}^{d_{1}}\otimes\mathbb{C}^{d_{2}}$ to $\left|b\right\rangle \otimes\left|a\right\rangle $
in $\mathbb{C}^{d_{2}}\otimes\mathbb{C}^{d_{1}}$.\tabularnewline
\hline 
$\mathbb{B}_{\left[d_{i},d_{j}\right]}$  & $\mathbb{B}_{d_{i}}\otimes\mathbb{B}_{d_{j}}$, tensor product of
the computational bases of $\mathbb{C}^{d_{i}}$ and $\mathbb{C}^{d_{j}}$.\tabularnewline
\hline 
$\left|ij\right\rangle _{\left[d_{1},d_{2}\right]}$  & An element of $\mathbb{B}_{\left[d_{i},d_{j}\right]}$, stands for
the state $\left|i\right\rangle _{d_{1}}\otimes\left|j\right\rangle _{d_{2}}$.\tabularnewline
\hline 
$_{\left[d_{i,},d_{j}\right]}\rho_{j}\left(\chi\right)$  & $d_{j}$-dimensional reduced density matrix corresponding to the second
subsystem, after tracing out $d_{i}-$dimensional first subsystem
from a state $\left|\chi\right\rangle $ in $\mathbb{C}^{d_{i}}\otimes\mathbb{C}^{d_{j}}$
tensor product space. \tabularnewline
\hline 
$_{\left[d_{i,},d_{j}\right]}\rho_{i}\left(\chi\right)$  & $d_{i}$-dimensional reduced density matrix corresponding to the first
subsystem, after tracing out $d_{j}-$dimensional second subsystem
from a state $\left|\chi\right\rangle $ in $\mathbb{C}^{d_{i}}\otimes\mathbb{C}^{d_{j}}$
tensor product space. \tabularnewline
\hline 
$S_{N}$  & Permutation group of $N-$symbols. \tabularnewline
\hline 
$\pi\left(d_{1},d_{2}\right)$  & Permutation corresponding to the permutation matrix $\hat{T}_{\left[d_{1},d_{2}\right]}$.
Element of the permutation $S_{N=d_{1}d_{2}}$.\tabularnewline
\hline 
$\mathbb{B}_{\left[d_{i},d_{j}\right]}^{T}$  & Set of eigenvectors of $\hat{T}_{\left[d_{1},d_{2}\right]}$, seen
as a basis for $\mathbb{C}^{N}$. Not related to $\mathbb{B}_{\left[d_{i},d_{j}\right]}$
(except through a unitary transformation). \tabularnewline
\hline 
$\mathbb{S}_{\left[d_{1},d_{2}\right]}^{\eta}$  & Eigenspace of $\hat{T}_{\left[d_{1},d_{2}\right]}$ corresponding
to eigenvalue $\eta$. $\mathbb{S}_{\left[d_{1},d_{2}\right]}^{1}$
is the symmetric subspace and $\mathbb{S}_{\left[d_{1},d_{2}\right]}^{-1}$
is the anti-symmetric subspace. \tabularnewline
\hline 
$R_{\left[d_{1},d_{2}\right]}$  & Completely entangled subspace in the $\mathbb{C}^{d_{1}}\otimes\mathbb{C}^{d_{2}}$
tensor product space according to Bhat. \tabularnewline
\hline 
\end{tabular}
\par\end{centering}

\caption{List of symbols relevant to the bipartite decomposition.\label{tab:Bipartite_Symbols}}
\end{table*}

\begin{table*}[p]
\begin{tabular}{|c|>{\raggedright}m{15cm}|}
\hline 
\textbf{Symbol}  & \centering{}\textbf{Description}\tabularnewline
\hline 
$\boldsymbol{d}=\left[d_{1},d_{2},\cdots,d_{k}\right]$  & A multiplication decomposition of $N$. Positive integers $>1$ such
that $\prod d_{i}=N$. \tabularnewline
\hline 
$\mathbb{P}\left(N\right)$  & All multiplicative partitions of $N$, $\left[1,N\right]$ and $\left[N,1\right]$
are not included in the definition.\tabularnewline
\hline 
$\mathbb{P}_{k}\left(N\right)$  & All multiplicative partitions of $N$ having $k$ terms.\tabularnewline
\hline 
$\mathbb{E}\left(\boldsymbol{d}\right)$  & Set of all partitions of $N$ which are connected to $\boldsymbol{d}$
by a permutation. \tabularnewline
\hline 
$k=n\left(\boldsymbol{d}\right)$  & Number of elements in $\boldsymbol{d}$. Number of subsystems in the
tensor product space $\mathbb{C}^{\boldsymbol{d}}$. \tabularnewline
\hline 
$\sigma$  & Appears along with $\boldsymbol{d}$. Refers to any permutation of
$k$ symbols, where $k=n\left(\boldsymbol{d}\right)$.\tabularnewline
\hline 
$\sigma\left(\boldsymbol{d}\right)$  & Shorthand notation for $\left[d_{\sigma^{-1}\left(1\right)},d_{\sigma^{-1}\left(2\right)},\cdots,d_{\sigma^{-1}\left(k\right)}\right]$. \tabularnewline
\hline 
$\mathbb{C}^{\boldsymbol{d}}$  & Tensor product space $\mathbb{C}^{d_{1}}\otimes\mathbb{C}^{d_{1}}\otimes\mathbb{\cdots\otimes C}^{d_{k}}$.\tabularnewline
\hline 
$\mathbb{C}^{\sigma\left(\boldsymbol{d}\right)}$  & Tensor product space $\mathbb{C}^{d_{\sigma^{-1}\left(1\right)}}\otimes\mathbb{C}^{d_{\sigma^{-1}\left(2\right)}}\otimes\cdots\mathbb{\otimes C}^{d_{\sigma^{-1}\left(k\right)}}$.\tabularnewline
\hline 
$\mathbb{B}_{\boldsymbol{d}}$  & Tensor product of the $k$ computational bases $\mathbb{B}_{d_{1}},\mathbb{B}_{d_{2}},\cdots,\mathbb{B}_{d_{k}}$in
that order. \tabularnewline
\hline 
$\left|i_{1}i_{2}\cdots i_{k}\right\rangle _{\mathbf{d}}$  & An element of $\mathbb{B}_{\boldsymbol{d}}$. Shorthand notation for
$\left|i_{1}\right\rangle \otimes\left|i_{2}\right\rangle \otimes\cdots\otimes\left|i_{k}\right\rangle $
where each $\left|i_{r}\right\rangle \in\mathbb{B}_{d_{r}}$.\tabularnewline
\hline 
$\hat{T}_{\mathbf{d},\sigma}$  & A mapping between states $\left|i_{1}i_{2}\cdots i_{k}\right\rangle _{\mathbf{d}}$
and $\left|\sigma\left(i_{1}\right)\sigma\left(i_{2}\right)\cdots\sigma\left(i_{k}\right)\right\rangle _{\sigma\left(\mathbf{d}\right)}$.\tabularnewline
\hline 
$\mathbb{S}_{\boldsymbol{d},\sigma}^{\eta}$  & Eigenspace of $\hat{T}_{\mathbf{d},\sigma}$ corresponding to an eigenvalue
$\eta$. $\mathbb{S}_{\boldsymbol{d},\sigma}^{1}$ is the symmetric
subspace and $\mathbb{S}_{\boldsymbol{d},\sigma}^{-1}$ represents
the anti-symmetric subspace. \tabularnewline
\hline 
$\pi\left(\mathbf{d},\sigma\right)$  & Permutation matrix corresponding to the permutation $\hat{T}_{\mathbf{d},\sigma}$.\tabularnewline
\hline 
$\Omega\left(N\right)$  & Number of prime factors of $N$, allowing for repetition. \tabularnewline
\hline 
$\mathbf{d}_{p}$  & A prime partition $\left[d_{1},d_{2},\cdots,d_{\Omega\left(N\right)}\right]$,
such that all $d_{i}$s are prime and $d_{i}\leq d_{j}$ if $i<j$.\tabularnewline
\hline 
$\boldsymbol{d}^{'}$  & A coarse-grained partition. $\boldsymbol{d}$ with $n\left(\boldsymbol{d}\right)<\Omega\left(N\right)$ \tabularnewline
\hline 
$\sigma^{'}$  & Appears along with $\boldsymbol{d}^{'}$. Permutation $\in S_{n\left(\boldsymbol{d}^{'}\right)}$. \tabularnewline
\hline 
$\sigma_{c}$  & Given along with a $\boldsymbol{d}$, refers to the cyclic shift of
subsystems, $\left(1,2,\cdots k\right)$ where $k=n\left(\boldsymbol{d}\right)$.\tabularnewline
\hline 
\end{tabular}

\caption{List of symbols relevant to multipartite decomposition.\label{tab:MP_Symbols}}
\end{table*}

 \bibliographystyle{apsrev4-1}
%

\end{document}